
%
%
%
%

\input amstex
\documentstyle{amsppt}
\topmatter
\author Thomas P. Branson${}^\dag$, Peter B. Gilkey${}^\ddag$, Dmitri V.
Vassilevich${}^*$\endauthor
\title The asymptotics of the Laplacian on a manifold with boundary II\endtitle
\thanks ${}^\dag$ Research partially supported by an
international travel grant of the NSF (USA)\endthanks
\thanks ${}^\ddag$ Research partially supported by the NSF
(USA) and IHES (France)\endthanks
\thanks ${}^*$ Research partially supported by GRACENAS (Russia) \endthanks
\keywords Laplacian, spectral geometry, heat equation asymptotics\endkeywords
\subjclass Primary 58G25\endsubjclass
\abstract We study the fifth term in the
asymptotic expansion of the heat operator trace for a partial differential
operator
of Laplace type on a compact Riemannian manifold with Dirichlet or Neumann
boundary conditions.\endabstract

\endtopmatter
\rightheadtext{The asymptotics of the Laplacian II}
\leftheadtext{T. Branson, P. Gilkey, and D. Vassilevich}
\def\refBe{1}\def\refBKE{2}\def\refBGa{3}\def\refBGb{4}
\def\refGa{5}\def\refGr{6}\def\refGu{7}\def\refKe{8}
\def\refKea{9}\def\refLe{10}\def\refSm{11}\def\refSW{12}\def\refVa{13}
\def\et{\smallbreak}
\def\PGa{\smallskip\noindent\qquad\qquad}
\def\PGb{\smallskip\noindent\qquad\qquad\quad}
\def\PGc{\smallskip\noindent\qquad\qquad\qquad}
\def\px(#1){\smallbreak\item{(#1)}}\newcount\qcta\global\qcta=1
\newcount\qctL\global\qctL=0
\def\pz{\global\advance\qctL by\qcta\item {\phantom{xx}\rm\number\qctL)}}
\def\ptr{\operatorname{Tr}}\def\cal{\Cal}
\def\ptrl{\operatorname{Tr}_{L^2}}\def\preal{\Bbb R}\def\pinteger{\Bbb Z}
\def\birdy{\textstyle{d\over{d\epsilon}}|\sb{\epsilon=0}}
\document
\head\S1 Statement of results\endhead
Let
$   M$ be a smooth compact Riemannian
manifold of dimension
$  m$ with smooth boundary
$  \partial M.$ Let
$  V$ be a smooth vector bundle over
$  M$ and let
$  D$ be a partial differential operator of Laplace type on
$ C \sp{  \infty }(V).$ There exists a unique
connection
$  \nabla $ on
$  V$ and a unique endomorphism
$  E$ of
$  V$ so that
$$
D=-(\ptr( \nabla  \sp{ 2})+E).
$$
If
$  \phi  \in C \sp{  \infty }(V),$ let
$  \phi  \sb{ ;m}$ be the covariant derivative
of
$  \phi $ with respect to the inward unit
normal. Let
$  S$ be an endomorphism of
$ V |  \sb{  \partial M}.$ We define the
modified Neumann boundary operator
$  {\cal B}  \sb{ S} \sp{ +}$ and the Dirichlet
boundary operator
$  {\cal B}  \sp{ -}$ by
$$
{\cal B}  \sb{ S} \sp{ +} \phi :=( \phi
\sb{ ;m}+S \phi ) |  \sb{  \partial M}~\text{\text{
and}}~ {\cal B}  \sp{ -} \phi := \phi  |
\sb{  \partial M}.
$$
Let
$  {\cal B} = {\cal B}  \sp{  \pm } \sb{
S}$ denote either boundary operator; we set
$ S=0$ for Dirichlet boundary conditions
to have a uniform notation. Let
$ D \sb{  {\cal B} }$ be the operator defined
by the appropriate boundary condition. If
$  F$ is a smooth function on
$  M,$ there is an asymptotic series as
$ t \downarrow 0$ of the form
$$
\ptrl(Fe \sp{ -tD\sb{ {\cal B} }}) \cong
\Sigma  \sb{ n \geq 0}t \sp{ (n-m)/2}a \sb{
n}(F,D, {\cal B} )
$$
where the
$ a \sb{ n}(F,D, {\cal B} )$ are locally
computable; see \cite{\refGr} for details.
We computed
$ a \sb{ n}$ for
$ n \leq 4$ in \cite{\refBGa}; we have changed
notation slightly from that paper. In this
paper, we compute
$ a \sb{ 5}$ if the boundary is totally geodesic.
If the boundary is not totally geodesic,
the second fundamental form
$  L$ enters and the combinatorial complexity
becomes formidable. In this setting, we also obtain a result
by restricting attention to regions in $\preal^m$. In particular, we
determine the coefficient of
$  |  \nabla L |  \sp{ 2}$ in
$ a \sb{ 5}( \Delta  \sb{ 0}, {\cal B}_0^\pm )$
for the Laplacian $\Delta_0$ on functions with pure
Neumann or Dirichlet boundary conditions.
In the Neumann case, this coefficient plays an important role
in the work of Gursky \cite{\refGu} on compactness of
isospectral sets in planar domains. The determination of
this coefficient was, in large part, the original motivation for the
present paper.

\par\rm
We adopt the following notational conventions.
Let indices
$  i,$
$  j,$ ... range from 1 through
$  m$ and index a local orthonormal frame
$  \{ e \sb{ 1},...,e \sb{ m} \} $ for the
tangent bundle of
$  M.$ On the boundary, let indices
$  a,$
$  b,...$ range from 1 through
$  m-1$ and index a local orthonormal
frame for the tangent bundle of the boundary;
let
$ e \sb{ m}$ be the inward unit normal. We
adopt the Einstein convention and sum over
repeated indices. Let
$ R \sb{ ij kl}$ be the
components of the curvature tensor of the
Levi-Civita connection on
$  M.$ With our sign convention,
$ R \sb{ 1212}$ is negative on the standard
sphere in Euclidean space. The Ricci tensor
$  \rho $ and the scalar curvature
$  \tau $ are given by
$$
\rho  \sb{ ij}:=R \sb{ ik kj}~\text{\text{
and}}~ \tau = \rho  \sb{ ii}=R \sb{ ik
ki}.
$$
Let
$  \rho  \sp{ 2}:= \rho  \sb{ ij} \rho  \sb{
ij}$ and
$ R \sp{ 2}:=R \sb{ ij kl}R \sb{ ij kl}$
be the norm of the Ricci and full curvature
tensors. Let
$  \Omega  \sb{ ij}$ be the endomorphism
valued components of the curvature of the
connection on
$  V$, and let
$ L \sb{ ab}$ be the components of the second
fundamental form;
$$
L \sb{ ab}:=( \nabla  \sb{ e\sb{a}}e\sb{b},e \sb{
m}).
$$
Let \lq ;\rq\ denote multiple covariant differentiation
with respect to the Levi-Civita connection
of
$  M$ and let \lq:\rq\ denote multiple
tangential covariant differentiation on the
boundary with respect to the Levi-Civita
connection of
$  \partial M;$ the difference between these
two is measured by the second fundamental
form. When sections of bundles built from
$  V$ are involved, \lq;\rq\ will mean
$  \nabla  \sp{ M} \otimes 1+1 \otimes  \nabla  \sp{
V}$ and \lq:\rq\ will mean
$  \nabla  \sp{  \partial M} \otimes 1+1 \otimes  \nabla  \sp{
V|_{\partial M}}.$ Thus
$ E \sb{ ;a}=E \sb{ :a}$ since there are
no tangential indices in
$  E$ to be differentiated, while
$ E \sb{ ;ab}$ and
$ E \sb{ :ab}$ do not agree since the index
$  a$ is also being differentiated. Since $L$ and $S$
are only defined on the boundary, these tensors can only
be differentiated tangentially. Let
$  dx$ and
$  dy$ be the Riemannian volume elements
on
$  M$ and on
$  \partial M.$ If
$ f \sb{ 1} \in C \sp{  \infty }(M)$ and
if
$ f \sb{ 2} \in C \sp{  \infty }( \partial M),$
define
$$
f \sb{ 1}[M]:= {\textstyle\int}  \sb{ M}f \sb{
1}(x)dx~\text{\text{and}}~f \sb{ 2}[ \partial
M]= {\textstyle\int}  \sb{  \partial M}f \sb{
2}(y)dy.
$$
\proclaim{Theorem 1.1} Suppose the boundary
of $M$ is totally geodesic. Then
\par\noindent
\rm \smallbreak\noindent\qquad
$  a \sb{ 5}(F,D, {\cal B}  \sb{ S} \sp{
\pm })= \pm 5760 \sp {-1} (4 \pi ) \sp{
(m-1)/2}\ptr \{ F(360E \sb{ ;mm}+1440E \sb{
;m}S+720E \sp{ 2 } +240E \sb{ :aa}+240 \tau E+120 \Omega  \sb{
ab} \Omega  \sb{ ab}+48 \tau  \sb{ ;ii}+20 \tau  \sp{
2}-8 \rho  \sp{ 2}+8R \sp{ 2 } -120 \rho  \sb{ mm}E-20 \rho  \sb{ mm} \tau
+480 \tau S \sp{ 2}+(90 \sp{ +},-360 \sp{ -}) \Omega  \sb{
am} \Omega  \sb{ am}+12 \tau  \sb{ ;mm} +24 \rho  \sb{ mm:aa}+15 \rho  \sb{
mm;mm}+270 \tau  \sb{ ;m}S+120 \rho  \sb{ mm}S \sp{ 2}+960S \sb{
:aa}S  +600S \sb{ :a}S \sb{ :a}+16R \sb{ am mb} \rho  \sb{
ab}-17 \rho  \sb{ mm} \rho  \sb{ mm}-10R \sb{
am mb}R \sb{ am mb } +2880ES \sp{ 2}+1440S \sp{ 4}~)+F \sb{
;m}((195/2 \sp{ +},60 \sp{ -})   \tau  \sb{
;m } +240 \tau S-90 \rho  \sb{ mm}S+270S \sb{
:aa}+$\hfil\break$
(630 \sp{ +},450 \sp{ -})E \sb{ ;m } +1440ES+720S \sp{ 3})+F \sb{
;mm}(60 \tau -90 \rho  \sb{ mm}+360E+360S \sp{ 2})  +180SF \sb{ ;m mm}+45F
\sb{ ;mm mm} \} [ \partial M].$\endproclaim\medbreak

\par\rm
As mentioned above, if
$  \partial M$ is not totally geodesic, the
number of invariants becomes unmanageable;
the general formula involves 63 additional
terms; see Lemma 5.1 for details. However, if we restrict to the Laplacian
$\Delta_0$
on functions with Neumann or Dirichlet boundary
conditions, assume
$  M$ is a domain in
$  \preal  \sp{ m},$ and set
$ F=1,$ a great simplification occurs. Recall
that
$ L \sb{ bc:a}-L \sb{ ac:b}=R \sb{ ab cm}$, see for example \cite{\refBGa,
Lemma
A.1 (b)}. Consequently, if the curvature tensor
$  R$ vanishes, then
$  \nabla L$ is a symmetric 3-tensor. Thus
there exist universal constants such that

\par\noindent
\PGa
$ a \sb{ 5}( \Delta  \sb{ 0}, {\cal B}_0  \sp{
\pm })= \pm (4 \pi )^{(m-1)/2}5760 \sp {-1}  {\textstyle\int}  \{ e \sp{
\pm } \sb{ 1}L \sb{ ab:c}L \sb{ ab:c}+e \sp{
\pm } \sb{ 2}L \sb{ aa}L \sb{ bb}L \sb{
cc}L \sb{ dd} $

\par\noindent
\PGb
$ +e \sp{  \pm } \sb{ 3}L \sb{ ab}L \sb{
ab}L \sb{ cc}L \sb{ dd}+e \sp{  \pm } \sb{
4}L \sb{ ab}L \sb{ ab}L \sb{ cd}L \sb{ cd}+e \sp{
\pm } \sb{ 5}L \sb{ ab}L \sb{ bc}L \sb{
ca}L \sb{ dd} $

\par\noindent
\PGb
$ +e \sp{  \pm } \sb{ 6}L \sb{ ab}L \sb{
bc}L \sb{ cd}L \sb{ da} \} [ \partial M].$
\et

\proclaim{Theorem 1.2}
$  e \sb{ 1} \sp{ +}=-19 \cdot 45/16$ \it
and
$ e \sb{ 1} \sp{ -}=45/16. $\endproclaim
\proclaim{Remark:} \rm The constant
$  e \sp{ +} \sb{ 1}$ is of particular interest
as it controls certain compactness estimates
for Neumann boundary conditions and plays
an important role in theorems of Gursky \cite{\refGu};
it was not previously known and its determination
was one of the primary goals of this study.
The value of
$ e \sb{ 1} \sp{ -}$ was previously known; see
\cite{\refSm,\refSW}. Levitin \cite{\refLe} has used
the algorithm of Kennedy \cite{\refKe, \refKea}
to compute
$ a \sb{ 5}$ for the ball in
$  \preal  \sp{ m}$ for any
$  m;$ this result has been used by van
den Berg \cite{\refBe} to determine
$ e \sb{ 2} \sp{  \pm },$
$ e \sb{ 3} \sp{  \pm },$
$ (e \sb{ 4}+e \sb{ 5}) \sp{  \pm },$ and
$ e \sb{ 6} \sp{  \pm }.$ See also related
work by Bordag et al \cite{\refBKE}.\endproclaim\medbreak

\par\rm
Here is a brief outline to the paper. In
\S2, we recall some functorial properties
of the invariants
$ a \sb{ n}( \cdot )$, and recall the calculation
of
$ a \sb{ n}$ for
$ n \leq 4.$ In \S3, we compute
$ a \sb{ 5}$ for the special case
$ m=1;$ this is an important step in the general case. In \S4, we prove Theorem
1.1; we shall omit the calculation of the coefficient
of
$  \Omega  \sb{ am} \Omega  \sb{ am}$ as
it is fairly lengthy; details are available
from the authors. In \S5, we prove Theorem
1.2. \S6 is an appendix in which we give some variational formulas used
elsewhere in the paper.\bigbreak
\head \S2 Functorial properties\endhead
We summarize below properties established
in \cite{\refBGa}:

\proclaim{Lemma 2.1}\global\qctL=0
\pz  Let
$  N \sp{  \nu }(F)=F \sb{ ;m ... m}$ be
the
$  \nu  \sp{ th}$ normal covariant derivative.
There exist invariant local formulae
$ a \sb{ n}(x,D)$ and
$ a \sb{ n, \nu }(y,D)$ so that:
$$
a \sb{ n}(F,D, {\cal B} )= \{ Fa \sb{ n}(x,D) \}
[M]+ \{  \Sigma  \sb{ 0 \leq  \nu  \leq
n-1}~N \sp{  \nu }(F)a \sb{ n, \nu }(y,D, {\cal B}
) \} [ \partial M].
$$
The interior invariant
$  a \sb{ n}(x,D)$ is homogeneous of order
$  n$ in the jets of the symbol of
$  D$ and vanishes for
$  n$ odd; it is independent of the boundary
condition chosen. The boundary invariants
$ a \sb{ n, \nu }$ are homogeneous of order
$ n- \nu -1. $
\pz Let
$ M=M \sb{ 1} \times M \sb{ 2}$ and
$ D=D \sb{ 1} \otimes 1+1 \otimes D \sb{
2}.$ Let
$  \partial M \sb{ 1}= \emptyset .$ Then
$$
a \sb{ n}(x,D)= \Sigma  \sb{ p+q=n}a \sb{
p}(x \sb{ 1},D \sb{ 1})a \sb{ q}(x \sb{
2},D \sb{ 2})
$$
$$
a \sb{ n, \nu }(y,D)= \Sigma  \sb{ p+q=n}a \sb{
p, \nu }(x \sb{ 1},D \sb{ 1})a \sb{ q}(y \sb{
2},D \sb{ 2}, {\cal B} ).
$$
\pz If we expand
$  a \sb{ n}(x,D)$ or
$ a \sb{ n, \nu }(y,D, {\cal B} )$ with respect
to a Weyl basis, the coefficients depend
on the dimension
$  m$ only through a normalizing constant.
They are independent of the dimension of
$  V. $\smallbreak
\pz If
$ D( \epsilon )=e \sp{ -2 \epsilon F}D,$
then
$  \birdy a \sb{ n}(1,D, {\cal B} )=(m-n)a \sb{
n}(F,D, {\cal B} ). $\smallbreak
\pz If
$ D( \epsilon )=D- \epsilon F \cdot I \sb{
V},$ then
$  \birdy a \sb{ n}(1,D, {\cal B} )=a \sb{
n-2}(F,D, {\cal B} ). $
\pz If
$ D( \epsilon )=D- \epsilon  \cdot I \sb{
V},$ then
$  \birdy a \sb{ n}(F,D, {\cal B} )=a \sb{
n-2}(F,D, {\cal B} ). $\smallbreak
\pz If
$ m=n+2,$ then
$  \birdy a \sb{ n}(e \sp{ -2 \epsilon f}F,e \sp{
-2 \epsilon f}D, {\cal B} )=0.$\smallbreak
\pz  Let
$ m=1$ and let
$ b \in C \sp{  \infty }[0,1]$ be real. Let
$ A= \partial  \sb{ x}-b,$
$ A \sp{ *}=- \partial  \sb{ x}-b,$
$ D \sb{ 1}=A \sp{ *}A,$
$ D \sb{ 2}=AA \sp{ *},$ and
$ S=b.$ Then
$$
{}~(n-1)(a \sb{ n}(F,D \sb{ 1}, {\cal B}
\sp{ -})-a \sb{ n}(F,D \sb{ 2}, {\cal B}
\sb{ S} \sp{ +}))=a \sb{ n-2}( \partial
\sb{ x} \sp{ 2}F+2b \partial  \sb{ x}F,D \sb{
1}, {\cal B}  \sp{ -}).
$$
\endproclaim
\proclaim{Remark} \rm  Assertion (3) is a bit
formal; it is best illustrated by reference
to Theorems 2.4 and 2.5 below which express
$ a \sb{ n}$ for
$ n \leq 4$ in terms of a Weyl basis.\endproclaim\medbreak

\par\rm
There are two additional properties which
we shall need which were not discussed in
\cite{\refBGa}. The first relates certain
coefficients in a Weyl basis for Neumann
boundary conditions to the corresponding
coefficients for Dirichlet boundary conditions.
Suppose that
$   \tilde M=N \times S \sp{ 1}$ and that
$  V$ is the trivial line bundle. Let
$ T(y, \theta )=(y,- \theta )$ define an
involution of
$   \tilde M$ where
$  \theta  \in  \preal /2 \pi  \pinteger =S \sp{
1}$ is the usual periodic parameter. Let
$ M=N \times [0, \pi ] \subset   \tilde M$
and let
$  F$ be a smooth function on
$   \tilde M$ which is preserved by
$  T.$ Let
$ e \sb{ m}$ be the inward unit normal on
$  M;$
$ e \sb{ m}= \partial  \sb{  \theta }$ at
$  \theta =0$ and
$ e \sb{ m}=- \partial  \sb{  \theta }$
at
$  \theta = \pi .$ We assume
$  D$ is preserved by
$  T;$ this means that if
$ D=- \{ g \sp{ ij} \partial  \sb{ i} \partial  \sb{
j}+A \sp{ k} \partial  \sb{ k}+B \} ,$ then

\par\noindent
\PGa
$ g \sp{ ab}(y, \theta )=g \sp{ ab}(y,- \theta ),$
$ g \sp{ mm}(y, \theta )=g \sp{ mm}(y,- \theta ), $

\par\noindent
\PGa
$ g \sp{ am}(y, \theta )=-g \sp{ am}(y,- \theta ),$
$ A \sp{ m}(y, \theta )=-A \sp{ m}(y,- \theta ), $

\par\noindent
\PGa
$ A \sp{ a}(y, \theta )=A \sp{ a}(y,- \theta ),$
$ B(y, \theta )=B(y,- \theta ).$ \smallbreak

\par\noindent
Let
$  {\cal B}  \sp{  \pm }$ denote pure Neumann
and Dirichlet boundary conditions so that
$ S=0. $

\proclaim{Lemma 2.2} We adopt the notation established above. Let $n$
be arbitrary and let $\nu$ be even. Then
$$
a \sb{ n, \nu }(y,D, {\cal B}  \sb{ 0} \sp{
+})+a \sb{ n, \nu }(y,D, {\cal B}  \sp{
-})=0.$$\endproclaim
\demo{Proof}  Let $F$ be an even function. Since
$  D$ is invariant under the involution
$  T,$ we may decompose the eigenfunctions of
$D$ on $\tilde M$ into the even and odd eigenfunctions.
The even eigenfunctions satisfy pure Neumann
boundary conditions on
$ M;$ the odd eigenfunctions satisfy
Dirichlet boundary conditions on
$ M.$ Thus
$$
Tr\sb{L\sp{2}M}(Fe \sp{ -tD\sb{-}})+Tr\sb{L\sp{2}M}(Fe \sp{
-tD\sb{+}})=Tr\sb{L\sp{2}(\tilde M)}(Fe \sp{ -tD}).
$$
We equate coefficients of
$  t$ in the asymptotic expansions of
both sides of this equation. The interior
integrals cancel since on the left hand side
we are integrating over
$ [0, \pi ]$ twice and on the right hand
side we are integrating over
$ [- \pi , \pi ].$ There are no boundary
integrals on the right hand side so the boundary
integrals on the left hand side must cancel. This shows
$$\Sigma_\nu\{ N^\nu(F)(a_{n,\nu}(y,D,{\cal B_0^+})+a_{n,\nu}(y,D,{\cal
B^-}))\}[\partial M]=0.$$ Since we may specify the $N^\nu(F)$ arbitrarily
for $\nu$ even, the desired vanishing theorem now follows.
\hfill\qed\enddemo\medbreak

\par\rm
In the next Lemma, we shall restrict to the case $n=5$ in the interests
of clarity and note that there is a more general principal applicable.

\proclaim{Lemma 2.3} When $a_5$ is expanded in a Weyl basis, the
coefficients of the following terms are zero:
$  F \sb{ ;m}\ptr( \Omega  \sb{ am;a}),$ $  F\ptr( \Omega  \sb{ am;ma}),$
$  F\ptr(S \Omega  \sb{ am;a}),$ $  F\ptr(S \sb{ :a} \Omega  \sb{ am}),$
$  FL \sb{ aa}\ptr( \Omega  \sb{ bm;b}). $
\endproclaim
\medbreak
\demo{Proof} We adapt an argument given in \cite{\refBGb}. Let $  F$ be real.
If
$  D$ and $  S$ are real, then $  \ptrl(Fe \sp{ -tD\sb{ {\cal B} }})$
is real so the coefficients of the terms listed above must be real.
On the other hand, if $  \nabla $ is unitary with respect to some
fiber metric on $  V$ and if $  E$ and
$  S$ are self-adjoint, then $ D \sb{  {\cal B} }$ is self-adjoint so
again $  \ptrl(Fe \sp{ -tD\sb{ {\cal B} }})$
is real. We take $  V$ to be a line bundle;
$  \Omega $ is pure imaginary in this context.
Thus the coefficients must be pure imaginary
as well and hence must vanish.\hfill\qed\enddemo\medbreak

\par\rm
We refer to \cite{\refBGa} for the computation
of
$ a \sb{ n}$ for
$ n \leq 4. $

\proclaim{Theorem 2.4 (Dirichlet)}\global\qctL=0
\pz $ a \sb{ 0}(F,D, {\cal B}  \sp{ -})=
   (4 \pi ) \sp{ -m/2}\ptr(F)[M].$
\pz
  $ a \sb{ 1}(F,D, {\cal B}  \sp{ -})= -4 \sp {-1} (4 \pi ) \sp{
  -(m-1)/2}\ptr(F)[ \partial M].$
\pz $ a \sb{ 2}(F,D, {\cal B}  \sp{ -})=
    (4 \pi ) \sp{ -m/2}6 \sp {-1}
    \{ \ptr(6FE+F \tau )[M]+\ptr(2FL \sb{
  aa}-3F \sb{ ;m})[ \partial M] \} $
\pz $ a \sb{ 3}(F,D, {\cal B}  \sp{ -})=
   -384 \sp {-1} (4 \pi ) \sp{ -(m-1)/2}
    \big\{ \ptr \{ F(96E+16 \tau -8 \rho
    \sb{ mm}+ 7L \sb{ aa}L \sb{ bb}-10L \sb{ ab}L \sb{
    ab})-30F \sb{ ;m}L \sb{ aa}+24F \sb{ ;mm} \} [ \partial M] \big\} $
\pz $ a \sb{ 4}(F,D, {\cal B}  \sp{ -})=
   (4 \pi ) \sp{ -m/2}360 \sp {-1}  \big\{ \ptr \{ F(60E \sb{
  ;kk}+60 \tau E+180E \sp{ 2}+30 \Omega  \sp{2}
  +12 \tau  \sb{ ;kk}+5 \tau  \sp{ 2}-2 \rho  \sp{
  2}+2R \sp{ 2}) \} [M]+\ptr \{ F(-120E \sb{;m}
  -18 \tau  \sb{ ;m}+120EL \sb{ aa}+20 \tau L \sb{
  aa}+4R \sb{ am am}L \sb{ bb}-12R \sb{
  am bm}L \sb{ ab}
  +4R \sb{ ab cb}L \sb{ ac}+24L \sb{
  aa:bb}+0L \sb{ ab:ab}+40/21L \sb{ aa}L \sb{
  bb}L \sb{ cc}+0 \Omega  \sb{ am;a}
    -88/7L \sb{ ab}L \sb{ ab}L \sb{ cc}+320/21L \sb{
  ab}L \sb{ bc}L \sb{ ac})+F \sb{ ;m}(-180E
  -30 \tau +0R \sb{ am am}-180/7L \sb{
  aa}L \sb{ bb}+60/7L \sb{ ab}L \sb{ ab})+
  24F \sb{ ;mm}L \sb{ aa}-30F \sb{ ;ii m}) \} [ \partial M] \big\}$
\endproclaim
\proclaim{Theorem 2.5 (Neumann)}\global\qctL=0
\pz
$ a \sb{ 0}(F,D, {\cal B}  \sb{ S} \sp{
+})=(4 \pi ) \sp{ -m/2}\ptr(F)[M]$.
\pz $ a \sb{ 1}(F,D, {\cal B}  \sb{ S} \sp{
+})=4 \sp {-1} (4 \pi ) \sp{ -(m-1)/2}\ptr(F)[ \partial M].$
\pz $ a \sb{ 2}(F,D, {\cal B}  \sp{ +})=
 (4 \pi ) \sp{ -m/2}6 \sp {-1}  \{ \ptr(6FE+F \tau )[M]+\ptr(2FL \sb{
aa}+3F \sb{ ;m} +12FS)$\hfil\break$[ \partial M] \}. $
\pz $ a \sb{ 3}(F,D, {\cal B}  \sb{ S} \sp{+})=
 384 \sp {-1} (4 \pi ) \sp{ -(m-1)/2} \big\{ \ptr \{ F(96E+16 \tau -8 \rho
\sb{ mm} +13L \sb{ aa}L \sb{ bb}+2L \sb{ ab}L \sb{
ab}+96SL \sb{ aa}+192S \sp{ 2})+F \sb{ ;m}(6L \sb{
aa}+96S)+24F \sb{ ;mm} \} [M] \big\}$
\pz $a \sb{ 4}(F,D, {\cal B}  \sb{ S} \sp{+})=
 (4 \pi ) \sp{ -m/2}360 \sp {-1}  \big\{ \ptr \{ F(60E \sb{
;kk}+60 \tau E+180E \sp{ 2}+30 \Omega  \sp{
2}+12 \tau  \sb{ ;kk}+5 \tau  \sp{ 2}-2 \rho  \sp{
2}+2R \sp{ 2}) \} [M]+\ptr \{ F(240E \sb{
;m}+42 \tau  \sb{ ;m}
 +24L \sb{ aa:bb}+0L \sb{ ab:ab}+120EL \sb{
aa}+20 \tau L \sb{ aa}+4R \sb{ am am}L \sb{
bb}-12R \sb{ am bm}L \sb{ ab}+4R \sb{
ab cb}L \sb{ ac}+40/3L \sb{ aa}L \sb{
bb}L \sb{ cc}+8L \sb{ ab}L \sb{ ab}L \sb{
cc}+32/3L \sb{ ab}L \sb{ bc}L \sb{ ac}+0 \Omega  \sb{
am;a}+720SE+120S \tau +0SR \sb{ am am}
 +144SL \sb{ aa}L \sb{ bb}+48SL \sb{ ab}L \sb{
ab}$\hfil\break$+480S \sp{ 2}L \sb{ aa}+480S \sp{ 3}+120S \sb{
:aa})+F \sb{ ;m}(180E+30 \tau +0R \sb{ am am}+12L \sb{
aa}L \sb{ bb}+12L \sb{ ab}L \sb{ ab}
 +72SL \sb{ aa}+240S \sp{ 2})+F \sb{ ;mm}(24L \sb{
aa}+120S)+30F \sb{ ;ii m} \} [ \partial M] \big\} $
\endproclaim
\head \S3 Dimension One\endhead

This section is devoted to the calculation
of
$ a \sb{ 5}$ in the one dimensional case;
this forms the foundation for our later computations.
We restrict to scalar operators; recall
$ S=0$ if
$  {\cal B} = {\cal B}  \sp{ -}. $
\proclaim{Theorem 3.1} $ a \sb{ 5}(F,D, {\cal B}  \sp{  \pm } \sb{
S})= \pm 384 \sp {-1}  \{ (24E \sb{ ;mm}+96E \sb{
;m}S+48E \sp{ 2}+192ES \sp{ 2}+96S \sp{
4})F  +((42 \sp{ +},30 \sp{ -})E \sb{ ;m}+96ES+48S \sp{
3})F \sb{ ;m}+(24E+24S \sp{ 2})F \sb{ ;mm }
 +12SF \sb{ ;m mm}+3F \sb{ ;mm mm} \} [ \partial M]. $
\endproclaim
\proclaim{Remark}\rm If we set
$ F=1,$ this agrees with Theorem 3.4 of \cite{\refGa}.\endproclaim
\demo{Proof} we use Lemma 2.1 (1) to
see there exist universal constants so that

\par\noindent
\PGa
$ a \sb{ 5}(F,D, {\cal B}  \sb{ S} \sp{
\pm })= \pm 384 \sp {-1}  \{ (b \sb{ 1} \sp{
\pm }E \sb{ ;mm}+b \sb{ 2} \sp{ +}E \sb{
;m}S+b \sb{ 3} \sp{  \pm }E \sp{ 2}+b \sb{
4} \sp{ +}ES \sp{ 2 }$

\par\noindent
\PGb
$ +b \sb{ 5} \sp{ +}S \sp{ 4})F+(b \sb{
6} \sp{  \pm }E \sb{ ;m}+b \sb{ 7} \sp{
+}ES+b \sb{ 8} \sp{ +}S \sp{ 3})F \sb{ ;m}+(b \sb{
9} \sp{  \pm }E+b \sb{ 10} \sp{ +}S \sp{
2})F \sb{ ;mm }$

\par\noindent
\PGb
$ +b \sb{ 11} \sp{ +}SF \sb{ ;m mm}+b \sb{
12} \sp{  \pm }F \sb{ ;mm mm} \} [ \partial M].$
\et

\par
Let $M$ be the interval $[0,\pi]$, and set $S=0$. Double
the interval to obtain a copy of the circle, and assume $E$ and $F$ are
even on the double. We apply Lemma 2.2 to see
$$
a \sb{ 5}(F,D, {\cal B}  \sb{ 0} \sp{ +})+a \sb{
5}(F,D, {\cal B}  \sp{ -})=0.
$$
This shows that
$ b \sb{ 1} \sp{ +}=b \sb{ 1} \sp{ -},$
$ b \sb{ 3} \sp{ +}=b \sb{ 3} \sp{ -},$
$ b \sb{ 9} \sp{ +}=b \sb{ 9} \sp{ -},$ and
$ b \sb{ 12} \sp{ +}=b \sb{ 12} \sp{ -}.$
No conclusion can be drawn about the terms
involving
$  S$ since we have set
$ S=0.$ Also, no conclusion can be drawn
about the term involving
$ E \sb{ ;m}$ since this will vanish for
an even function.

This shows that the general formula for $a_5$ in the one-dimensional case is

\par\noindent
\PGa
$ a \sb{ 5}(F,D, {\cal B}  \sb{ S} \sp{
\pm })= \pm 384 \sp {-1}  \{ (b \sb{ 1}E \sb{
;mm}+b \sp{ +} \sb{ 2}E \sb{ ;m}S+b \sb{
3}E \sp{ 2}+b \sp{ +} \sb{ 4}ES \sp{ 2 }$

\par\noindent
\PGb
$ +b \sb{ 5} \sp{ +}S \sp{ 4})F+(b \sb{
6} \sp{  \pm }E \sb{ ;m}+b \sb{ 7} \sp{
+}ES+b \sb{ 8} \sp{ +}S \sp{ 3})F \sb{ ;m}+(b \sb{
9}E+b \sb{ 10} \sp{ +}S \sp{ 2})F \sb{ ;mm }$

\par\noindent
\PGb
$ +b \sb{ 11} \sp{ +}SF \sb{ ;m mm}+b \sb{
12}F \sb{ ;mm mm} \} [ \partial M].$ \et

\par\noindent
By Theorems 2.4 and 2.5,

\par\noindent
\PGa
$ a \sb{ 3}(F,D, {\cal B}  \sb{ S} \sp{
\pm })= \pm 384 \sp {-1}  \{ (96E+192S \sp{
2})F+96SF \sb{ ;m}+24F \sb{ ;mm} \} [ \partial M].$
\et

\par\noindent
Let
$ D( \epsilon )=D- \epsilon I.$ Then
$ E( \epsilon )=E+ \epsilon I.$ By Lemma
2.1 (6)

\par\noindent
\PGa
$  \birdy a \sb{ 5}(F,D, {\cal B}  \sb{
S} \sp{  \pm })=a \sb{ 3}(F,D, {\cal B}  \sb{
S} \sp{  \pm }).$ \et

\par\noindent
This shows
$ b \sb{ 3}=48,$
$ b \sb{ 4} \sp{ +}=192,$
$ b \sb{ 7} \sp{ +}=96,$ and
$ b \sb{ 9}=24.$ Next let
$ D( \epsilon )=D- \epsilon F.$ Then
$ E( \epsilon )=E+ \epsilon F.$ By Lemma
2.1 (5)

\par\noindent
\PGa
$  \birdy a \sb{ 5}(1,D, {\cal B}  \sb{
S} \sp{  \pm })=a \sb{ 3}(F,D, {\cal B}  \sb{
S} \sp{  \pm }).$ \et

\par\noindent
It now follows that
$ b \sb{ 1}=24$ and
$ b \sb{ 2} \sp{ +}=96.$ This shows that

\par\noindent
\PGa
$ a \sb{ 5}(F,D, {\cal B}  \sb{ S} \sp{
\pm })= \pm 384 \sp {-1}  \{ (24E \sb{
;mm}+96E \sb{ ;m}S+48E \sp{ 2}+192ES \sp{
2}+b \sb{ 5} \sp{ +}S \sp{ 4})F $

\par\noindent
\PGb
$ +(b \sb{ 6} \sp{  \pm }E \sb{ ;m}+96ES+b \sb{
8} \sp{ +}S \sp{ 3})F \sb{ ;m}+(24E+b \sb{
10} \sp{ +}S \sp{ 2})F \sb{ ;mm }$

\par\noindent
\PGb
$ +b \sb{ 11} \sp{ +}SF \sb{ ;m mm}+b \sb{
12}F \sb{ ;mm mm} \} [ \partial M].$ \et

\par\rm
We now adopt the notation of Lemma 2.1 (8). Let $M=[0,1]$ and
let $b\in C^\infty[0,1]$ be real. Let
$ A= \partial  \sb{ x}-b,$
$ D \sb{ 1}=A \sp{ *}A,$ and
$ D \sb{ 2}=AA \sp{ *}.$ Then
$ E \sb{ 1}=-b \sb{ x}-b \sp{ 2}$ and
$ E \sb{ 2}=b \sb{ x}-b \sp{ 2}.$ We take
Dirichlet boundary conditions for
$ D \sb{ 1}$ and modified Neumann boundary
conditions given by
$ A \sp{ *}$ for
$ D \sb{ 2};$ this means
$ S=b.$  We assume
$  F$ vanishes to infinite order at the
endpoint
$ x=1.$ We expand
$ a \sb{ 3}$ with Dirichlet boundary conditions,

\par\noindent
\PGa
$ a \sb{ 3}(F \sb{ xx}+2bF \sb{ x},D \sb{
1}, {\cal B}  \sp{ -}) $

\par\noindent
\PGb
$ =-384 \sp {-1}  \{ 96(-b \sb{ x}-b \sp{
2})(F \sb{ xx}+2bF \sb{ x})+24(F \sb{ xx}+2bF \sb{
x}) \sb{ xx} \} (0) $

\par\noindent
\PGb
$ =-384 \sp {-1}  \{ (-96b \sb{ x}F \sb{
xx}-96b \sp{ 2}F \sb{ xx}-192b \sb{ x}bF \sb{
x}-192b \sp{ 3}F \sb{ x }$

\par\noindent
\PGc
$ +24F \sb{ xx xx}+48b \sb{ xx}F \sb{
x}+96b \sb{ x}F \sb{ xx}+48bF \sb{ xx x} \} (0) $

\par\noindent
\PGb
$ =-384 \sp {-1}  \{ 24F \sb{ xx xx}+48bF \sb{
xx x}-96b \sp{ 2}F \sb{ xx }$

\par\noindent
\PGc
$ +(48b \sb{ xx}-192b \sb{ x}b-192b \sp{
3})F \sb{ x} \} (0),$ \et

\par\noindent
$a \sb{ 5}$ with Neumann boundary conditions,

\par\noindent
\PGa
$ -a \sb{ 5}(F,D \sb{ 2}, {\cal B}  \sb{
S} \sp{ +})=-384 \sp {-1}  \{ (24(b \sb{
x}-b \sp{ 2}) \sb{ xx}+96(b \sb{ x}-b \sp{
2}) \sb{ x}b $

\par\noindent
\PGc
$ +48(b \sb{ x}-b \sp{ 2}) \sp{ 2}+192(b \sb{
x}-b \sp{ 2})b \sp{ 2}+b \sb{ 5} \sp{ +}b \sp{
4})F $

\par\noindent
\PGc
$ +(b \sb{ 6} \sp{ +}(b \sb{ x}-b \sp{ 2}) \sb{
x}+96(b \sb{ x}-b \sp{ 2})b+b \sb{ 8} \sp{
+}b \sp{ 3})F \sb{ x }$

\par\noindent
\PGc
$ +(24(b \sb{ x}-b \sp{ 2})+b \sb{ 10} \sp{
+}b \sp{ 2})F \sb{ xx}+b \sb{ 11} \sp{ +}bF \sb{
x xx}+b \sb{ 12}F \sb{ xx xx} \} (0),$
\et

\par\noindent
and
$ a \sb{ 5}$ with Dirichlet boundary conditions,

\par\noindent
\PGa
$ a \sb{ 5}(F,D, {\cal B}  \sp{ -})=-384 \sp {-1}  \{ (24(-b \sb{
x}-b \sp{ 2}) \sb{ xx}+48(-b \sb{ x}-b \sp{
2}) \sp{ 2})F $

\par\noindent
\PGc
$ +b \sb{ 6} \sp{ -}(-b \sb{ x}-b \sp{ 2}) \sb{
x}F \sb{ x}+24(-b \sb{ x}-b \sp{ 2})F \sb{
xx}+b \sb{ 12}F \sb{ xx xx} \} (0).$ \et

\par\noindent
By Lemma 2.1 (8),

\par\noindent
\PGa
$ a \sb{ 3}(F \sb{ xx}+2bF \sb{ x},D \sb{
1}, {\cal B}  \sp{ -})=4 \{ a \sb{ 5}(F,D \sb{
1}, {\cal B}  \sp{ -})-a \sb{ 5}(F,D \sb{
2}, {\cal B}  \sb{ S} \sp{ +}) \} .$ \et
\noindent
This leads to the equations\et
\centerline{\vbox{\offinterlineskip\halign{\vphantom{${}\sb{A}$}\vrule
 height 12pt\ #\hfil\ \vrule height
 12pt\ & #\hfil\ \vrule height 12pt\ \vrule height 12pt\ &
 #\hfil\ \vrule height 12pt\ & #\hfil\ \vrule height 12pt\cr
 \noalign{\hrule} Term & Equation & Term & Equation \cr\noalign{\hrule}
  $ b \sp{ 4}F $ &
  $ 0=48-192+b \sb{ 5} \sp{ +}+48$ &
  $ b \sp{ 3}F \sb{ x}$ &
  $ -192=4(-96+b \sb{ 8} \sp{ +})$\cr\noalign{\hrule}
  $ b \sb{ x}bF \sb{ x} $&
  $ -192=4(-2b \sb{ 6} \sp{ +}+96-2b \sb{6} \sp{ -})$&
  $ b \sb{ xx}F \sb{ x} $&
  $ 48=4(b \sb{ 6} \sp{ +}-b \sb{ 6} \sp{-})$\cr\noalign{\hrule}
  $ b \sp{ 2}F \sb{ xx} $&
  $ -96=4(-24+b \sb{ 10} \sp{ +}-24)$ &
  $ bF \sb{ x xx} $&
  $ 48=4b_{11}^+$\cr\noalign{\hrule}
  $ F \sb{ xx xx}$ &
  $ 24=4(b \sb{ 12}+b \sb{ 12})$&& \cr\noalign{\hrule}}}}
We solve this system of equations and complete the proof
by computing:\et
  \centerline{\vbox{\offinterlineskip\halign{
   \vphantom{${}\sb{A}$}#\hfil&\ #\hfil&\ #\hfil&#\hfil\cr
   $ b \sb{ 5} \sp{ +}=96,$ &$ b \sb{ 6} \sp{ +}=42,$ &
   $ b \sb{ 6} \sp{ -}=30,$ &$ b \sb{ 8} \sp{ +}=48,$\cr
   $ b \sb{ 10} \sp{ +}=24,$ & $ b \sb{ 11}^+=12,$ & $ b \sb{ 12}=3.$
   & $  \qed $\cr}}}\enddemo
\head \S4 Totally geodesic boundary\endhead

\par\rm
This section is devoted to the proof of Theorem
1.1. We assume for the remainder of this
section that the boundary of
$  M$ is totally geodesic. We begin the
proof of Theorem 1.1 by expressing
$ a \sb{ 5}$ in terms of universal expressions
with undetermined coefficients. Note that
$ 5760=15 \cdot 384=16 \cdot 360.$
\proclaim{Lemma 4.1}Suppose the boundary of
$   M$ is totally geodesic. There exist
universal constants so that
$  a \sb{ 5}(F,D, {\cal B}  \sp{  \pm } \sb{
S})= \pm (4 \pi ) \sp{ (m-1)/2}5760 \sp {-1} \ptr \{
 F(360E \sb{ ;mm}+1440E \sb{ ;m}S+720E \sp{
2}+2880ES \sp{ 2}+1440S \sp{ 4 }
 +240E \sb{ :aa}+240 \tau E+120 \Omega  \sb{
ab} \Omega  \sb{ ab}+48 \tau  \sb{ ;ii}+20 \tau  \sp{
2}-8 \rho  \sp{ 2}+8R \sp{ 2 }
 -120 \rho  \sb{ mm}E-20 \rho  \sb{ mm} \tau +480 \tau S \sp{
2}+c \sb{ 1} \sp{  \pm } \Omega  \sb{ am} \Omega  \sb{
am }
 +c \sb{ 2} \tau  \sb{ ;mm}+c \sb{ 3} \rho  \sb{
mm:aa}+c \sb{ 4} \rho  \sb{ mm;mm}+c \sb{
5} \sp{ +} \tau  \sb{ ;m}S
 +c \sb{ 6} \sp{ +} \rho  \sb{ mm}S \sp{
2}+c \sb{ 7} \sp{ +}S \sb{ :aa}S+c \sb{
8} \sp{ +}S \sb{ :a}S \sb{ :a }
+c \sb{ 9}R \sb{ am mb} \rho  \sb{
ab}+c \sb{ 10} \rho  \sb{ mm} \rho  \sb{
mm}+c \sb{ 11}R \sb{ am mb}R \sb{ am mb }
 +F \sb{ ;m}((630 \sp{ +},450 \sp{ -})E \sb{
;m}+1440ES+720S \sp{ 3 }
 +c \sb{ 12} \sp{  \pm } \tau  \sb{ ;m}+240 \tau S+c \sb{
13} \sp{ +} \rho  \sb{ mm}S+c \sb{ 14} \sp{
+}S \sb{ :aa})+F \sb{ ;mm}(360E+360S \sp{ 2}+60 \tau +c \sb{
15} \rho  \sb{ mm})
 +180SF \sb{ ;m mm}+$\hfil\break$45F \sb{ ;mm mm} \}
[\partial M].$
\endproclaim
\demo{Proof} Since the boundary is totally
geodesic, we may interchange \lq;\rq\ and
\lq:\rq\ as convenient. We write down a basis
for the space of invariants using H\. Weyl's
theorem. The only tricky part of this analysis
is the study of terms of order 4 in the jets of the
metric; these terms are studied at the end
of the proof. We omit the curvature terms
$ R \sb{ ma bc}R \sb{ ma bc},$ and
$  \rho  \sb{ mc} \rho  \sb{ mc}$ since these
vanish on a totally geodesic boundary. Since
$  \tau  \sb{ ;m}=2 \rho  \sb{ mm;m},$ we
also omit the terms
$  \rho  \sb{ mm;m}S$ and
$ F \sb{ ;m} \rho  \sb{ mm;m}$ from our list.
Finally, we use Lemma 2.3 to omit the following
invariants from our list:

\par\noindent
\PGa
$  \ptr(F \Omega  \sb{ am;am}),$
$  \ptr(F \Omega  \sb{ am;a}S),$
$  \ptr(F \Omega  \sb{ am}S \sb{ ;a}),$
$  \ptr(F \sb{ ;m} \Omega  \sb{ am;a}).$
\et

\par\rm
We can use previous results to determine
many of the coefficients in a Weyl spanning
set:

\par\noindent
\item{1.} We use Theorem 3.1 to determine
the coefficients of the following expressions
which do not involve
$  \Omega ,$
$  R,$ or the tangential derivatives of
$  S$ and
$  E: $

\par\noindent
\PGa
$ FE \sb{ ;mm},$
$ FE \sb{ ;m}S,$
$ FE \sp{ 2},$
$ F ES \sp{ 2},$
$ FS \sp{ 4},$
$ F \sb{ ;m}E \sb{ ;m},$
$ F \sb{ ;m}ES,$

\par\noindent
\PGa
$ F \sb{ ;m}S \sp{ 3},$
$ F \sb{ ;mm}E,$
$ F \sb{ ;mm}S \sp{ 2},$
$ F \sb{ ;mm m}S,$
$ F \sb{ ;mm mm}.$ \et

\par\noindent
\item{2.} We use Lemma 2.1 (6) to see
$  \birdy a \sb{ 5}(1,D- \epsilon I, {\cal B} )=a \sb{
3}(1,D, {\cal B} );$ this enables us to determine
the coefficient of
$  \tau E$ and
$  \rho  \sb{ mm}E$ in
$ a \sb{ 5}$ from the expression for
$ a \sb{ 3}$ given in Theorems 2.4 and 2.5.

\par\noindent
\item{3.} We apply Lemma 2.1 (2) to a product
$ M=M \sb{ 1} \times M \sb{ 2}$ where
$  \partial M \sb{ 1}= \emptyset $ to see

\par\noindent
\PGa
$ a \sb{ 5, \nu }(x \sb{ 1},y \sb{ 2},D \sb{
1} \otimes 1+1 \otimes D \sb{ 2}, {\cal B}  \sb{
S} \sp{  \pm })= \Sigma  \sb{ p+q=5}a \sb{
p }(x \sb{ 1},D \sb{ 1})a \sb{ q,\nu}(y \sb{
2},D \sb{ 2}, {\cal B}  \sb{ S} \sp{  \pm }).$
\et

\par\noindent
\itemitem{(i)} We take
$ (m \sb{ 1},m \sb{ 2})=(4,1)$ and study the terms
in $a_{5,0}$ which are independent of $M_2$.
These terms arise when $(p,q)=(4,1)$ so
$$\eqalign{a_{5,0}(\cdot,{\cal
B}_S^\pm)=\pm&(4\pi)^{(m-1)/2}\{60E^{D_1}_{:aa}+30(\Omega^{D_1})^2+
12\tau^{M_1}_{:aa}\cr
&+5(\tau^{M_1})^2-2(\rho^{M_1})^2+2(R^{M_1})^2+...\}/(4\cdot
384).\cr}$$ This determines the
coefficients of the invariants
$ E \sb{ :aa},$
$  \Omega  \sb{ ab} \Omega  \sb{ ab},$
$  \tau  \sb{ ;ii},$
$  \tau  \sp{ 2},$
$  \rho  \sp{ 2},$ and
$ R \sp{ 2}$ in $a_{5,0}$; still to be determined, of course,
is the coefficient of $\tau_{;mm}$. Similarly, we study
the cross terms to evaluate the coefficient of $\tau S^2$ in $a_{5,0}$, to
evaluate the coefficient
of $\tau S$ in $a_{5,1}$, and to evaluate the coefficient of $\tau$ in
$a_{5,2}$;
the cross terms arise when $(p,q)=(2,3)$. \et

\par\noindent
\itemitem{(ii)} We take
$ (m \sb{ 1},m \sb{ 2})=(2,2)$
and compute the cross terms involving the curvature tensor in $a_{5,0}$. These
arise when $(p,q)=(2,3)$ so
$$a_{5,0}(\cdot,{\cal B}_S^\pm)=\pm(4\pi)^{(m-1)/2}\tau^{M_1}(16\tau^{M_2}
-8\rho^{M_2}_{mm})/(6\cdot 384)+....$$
Since
$\tau^2=2\tau^{M_1}\tau^{M_2}+...$ and
$\tau\rho_{mm}=\tau^{M_1}\rho^{M_2}_{mm}+...$, the coefficient of
$\tau^2$ in $a_{5,0}$ is $16\cdot 5760/(2\cdot 6\cdot 384)=20$
and the coefficient
of $\tau\rho_{mm}$ in $a_{5,0}$ is $-8\cdot 5760/(6\cdot 384)=-20$.\et

\par\rm
We use Lemma 2.2 to see that
$ c \sb{  \mu } \sp{ +}=c \sb{  \mu } \sp{
-}$ with our normalizing sign convention
for many of the remaining coefficients; a
notable exception is the coefficient of
$  \Omega  \sb{ am} \Omega  \sb{ am}$ since
this term vanishes under the hypothesis of
Lemma 2.2.

\par\rm
We conclude the proof of Lemma 3.1 by studying
the terms which are linear in the second
covariant derivative
$  \nabla  \sp{ 2}R$ of the curvature  tensor.
We must set some of the indices equal to
$  m$ and contract the remaining indices
in pairs in an expression
$ R \sb{ i \sb{1} i \sb{2} i \sb{3} i \sb{4} ;i \sb{5} i \sb{6} }.$
If none of the indices
$ i \sb{  \mu }$ are equal to
$  m,$ we must contract 3 pairs of indices.
The usual argument then shows this result
is a multiple of
$  \tau  \sb{ ;ii}.$ Thus we can restrict
our attention to the case where either 2
or 4 of the indices
$ i \sb{  \mu }$ are equal to
$  m.$ If 4 of the indices are equal to
$  m,$ we get a multiple of
$  \rho  \sb{ mm;mm}.$ In the remaining case, 2 of the indices are
$  m$ and modulo lower order terms we
must consider the invariants
$$
R \sb{ ab ba;mm},~R \sb{ am ma;bb},
{}~R \sb{ ab am;bm},\ \text{and}\ R \sb{ am bm;ab}.$$
By the second Bianchi identity,
$
0=R \sb{ am bm;ab}+R \sb{ am ma;bb}+R \sb{
am ab;mb };
$
this eliminates
$ R \sb{ am bm;ab}$ from our list. Modulo
lower order terms, we may replace
$ R \sb{ am ab;mb}$ by
$ R \sb{ am ab;bm}.$ By the second Bianchi identity,
$$
0=R \sb{ ab am;bm}+R \sb{ ab mb;am}+R \sb{
ab ba;mm}=2R \sb{ ab am;bm}+R \sb{
ab ba;mm };
$$
this eliminate
$ R \sb{ ab am;bm}$ from our list as well.
We then note
$$
\eqalign{&\text{span} \{ R \sb{ ab ba;mm},
{}~R \sb{ am ma;bb},~ \tau  \sb{ ;ii},~ \tau
\sb{ ;mm} \} \cr~=&\text{span} \{  \tau
\sb{ ;ii},~ \tau  \sb{ ;mm},~ \rho  \sb{
mm:aa},~ \rho  \sb{ mm;mm} \} .~ \qquad\qed
\cr}
$$\enddemo

\par\rm
We complete the proof of Theorem 1.1 by evaluating
the constants which appear in Lemma 4.1.
We will not compute
$ c \sb{ 1} \sp{  \pm }$ as this requires
somewhat different techniques; details are
available from the authors upon request.
This term vanishes for the scalar Laplacian
in any event.

\par\rm
Let
$ D( \epsilon )=e \sp{ -2 \epsilon f}D,$
$ S( \epsilon )=e \sp{ - \epsilon f}S,$
$ F( \epsilon )=e \sp{ -2 \epsilon f}F.$
We assume
$ f \sb{ ;m} |  \sb{  \partial M}=0$ to preserve
the condition that the boundary is totally
geodesic. If
$ m=7,$ we use Lemma 2.1 (7) to see
$$
\birdy a \sb{ 5}(F,D)=0.
$$
We use Lemma 4.1 and results from \S6 to
compute
$  \birdy a \sb{ 5}(F,D).$ We integrate by
parts and combine terms to express
$  \birdy a \sb{ 5}(F,D)$ in terms of independent
integrands and derive the system of equations
\def\borko{\ \vrule height 12pt\ \vrule height 12pt\ }
\def\borka{\ \vrule height 12pt\ }
\et\centerline{\vbox{\offinterlineskip\halign{\vphantom{${}\sb{A}$}\vrule
 height 12pt\ #\hfil\ \vrule height 12pt\ &#\hfil&
 #\hfil& #\hfil\ \vrule height 12pt\cr
 \noalign{\hrule} Term & Equation & Term & Equation \cr\noalign{\hrule}
  $ Ff \sb{ ;mm mm} $&
  $39=2c \sb{ 2}+c \sb{ 4}$&\borko
  $ Ff \sb{ ;mm}S \sp{ 2}$&\borka
  $ 120=c \sb{ 6} \sp{ +}$\cr\noalign{\hrule}
  $ Ff \sb{ ;mm aa} $&
  $303=6c \sb{ 3}+12c \sb{ 2}+c \sb{4}$&\borko
  $ Ff \sb{ ;mm} \tau $&\borka
  $40=2c \sb{ 2}+c \sb{9}$\cr\noalign{\hrule}
  $ Ff \sb{ ;aa} \rho  \sb{ mm} $&
  $ 60=-2c \sb{ 3}-c \sb{ 9}-2c \sb{ 10}-c \sb{15}$&\borko
  $ Ff \sb{ ;a} \tau  \sb{ ;a} $&\borka
  $12=c \sb{ 2}$\cr\noalign{\hrule}
  $ Ff \sb{ ;a} \rho  \sb{ mm;a}$&
  $ 324=-c \sb{15}+12c \sb{ 2}+6c \sb{ 4}$&\borko
  $F \sb{ ;m}f \sb{ ;a}S \sb{ :a} $&\borka
  $270=c \sb{ 14} \sp{ +}$\cr\noalign{\hrule}
  $ Ff \sb{ ;a}R \sb{ am mb;b} $&
  $63=4c_2+c_4$&\borko
  $ Ff \sb{ ;ab} \rho  \sb{ ab} $&\borka
  $ 16=c \sb{ 9}$ \cr\noalign{\hrule}
  $F \sb{ ;m}f \sb{ ;mm m}$ &
  $(195\sp{ +},120 \sp{ -})=2c \sb{12} \sp{  \pm }$ &\borko
  $F \sb{ ;mm}f \sb{ ;aa} $&\borka
  $90=-c \sb{ 15}$\cr\noalign{\hrule}
  $ Ff \sb{ ;aa}S \sp{ 2}$&
  $ 1080=c \sb{ 6} \sp{ +}+c \sb{ 7} \sp{+}$&\borko
  $F \sb{ ;m}f \sb{ ;mm}S $&\borka
  $-90=c \sb{ 13} \sp{ +}$\cr\noalign{\hrule}
  $ Ff \sb{ ;a}S \sb{ :a}S $&
  $ 360=c \sb{ 7} \sp{ +}-c \sb{ 8} \sp{+}$&\borko
  $Ff \sb{ ;mm m}S $&\borka
  $270=c \sb{ 5} \sp{ +}$
  \cr\noalign{\hrule}
  $ Ff \sb{ ;ab}R \sb{ am mb}$&
  $378=5c_9+2c_{11}+24c_2+2c$&${}_4$&\cr\noalign{\hrule}
  $ Ff \sb{ ;mm} \rho  \sb{ mm} $&
  $346=-c_{15}-6c_{10}-c_{11}+12$&$c_2$&
 \cr\noalign{\hrule}}}}
   We solve this system of equations and complete the proof
   by computing:\et\centerline{\vbox{\offinterlineskip\halign{
   \vphantom{${}\sb{A}$}#\hfil&\ #\hfil&\ #\hfil&\ #\hfil&#\hfil\cr
   $ c \sb{ 2}=12,$ & $ c \sb{ 3}=24,$ &
   $ c \sb{ 4}=15,$ & $ c \sb{ 5} \sp{ +}=270,$ & $ c \sb{ 6} \sp{ +}=120,$\cr
   $ c \sb{ 7} \sp{ +}=960 $, & $ c \sb{ 8} \sp{ +}=600,$&
   $ c \sb{ 9}=16,$ & $ c \sb{ 10}=-17,$ & $ c \sb{ 11}=-10,$\cr
   $ c \sb{ 12}\sp+=195/2$, & $c_{12}^-=60$, &
   $ c \sb{ 13}\sp{ +}=-90,$ & $ c \sb{ 14} \sp{ +}=270,$ &
   $ c \sb{ 15}=-90. \quad\qed$\cr}}}
\head \S5 Domains in flat space\endhead

\par\rm
In this section, we complete the proof of
Theorem 1.2. We drop the assumption that the boundary of $M$ is totally
geodesic;
thus ``:'' and ``;'' differ. Lemma 4.1 generalizes to this context to become
\proclaim{Lemma 5.1} Suppose the boundary of $M$ is smooth
but not necessarily totally geodesic.
\item{\rm 1)} There exist universal
constants so that\medbreak
$a_{5}(F,D,{\cal B}_{S}^{\pm })
=\pm 5760^{-1}(4\pi)^{(m-1)/2}
\ptr\bigl\{F\{360E_{;mm}+1440E_{
;m}S+720E^{2}+$\hfil\break$240E_{:aa}+240\tau
E+120
\Omega
_{ab}\Omega_{ab}+48\tau _{;ii}+20\tau ^{
2}-8\rho^{2}+8R^{2}-120\rho_{mm}E-20\rho_{mm}\tau
+480\tau S^{2}+(90^{+},-360^{-})\Omega_{
am}\Omega_{am}+12\tau _{;mm}+24\rho_{mm:aa}+15\rho_{
mm;mm}+270\tau _{;m}S+120\rho_{mm}S^{2}+960S_{
:aa}S+600S_{:a}S_{:a}+16R_{ammb}\rho_{
ab}-17\rho_{mm}\rho_{mm}-10R_{
ammb}R_{ammb}+2880ES^{2}+1440S^{4}\}+F_{
;m}\{(195/2^{+},60^{-})\tau _{
;m}+240\tau S-90\rho_{mm}S+270S_{
:aa}+$\hfil\break$
(630^{+},450^{-})E_{;m}+1440ES+720S^{
3}\}+F_{;mm}\{60\tau -90\rho_{mm}+360E+360S^{
2}\}+180SF_{;mmm}+45F
_{;mmmm}+
F\{d_1^\pm L_{aa}E_{;m}
+d_2^\pm L_{aa}\tau _{:m}
+d_3^\pm L_{ab}R_{ammb;m}
+d_4^+L_{aa}S_{:bb}
+d_5^+L_{ab}S_{:ab}
+d_6^+L_{aa:b}S_{:b}
+d_7^+L_{ab:a}S_{:b}
+d_8^+L_{aa:bb}S
+d_9^+L_{ab:ab}S
+d_{10}^\pm L_{aa:b}L_{cc:b}
+d_{11}^\pm L_{ab:a}L_{cc:b}
+d_{12}^\pm L_{ab:a}L_{bc:c}
+d_{13}^\pm L_{ab:c}L_{ab:c}
+d_{14}^\pm L_{ab:c}L_{ac:b}
+d_{15}^\pm L_{aa:bb}L_{cc}
+$\hfil\break$d_{16}^\pm L_{ab:ab}L_{cc}
+d_{17}^\pm L_{ab:ac}L_{bc}
+d_{18}^\pm L_{aa:bc}L_{bc}
+d_{19}^\pm L_{bc:aa}L_{bc}
+1440^+L_{aa}SE
+$\hfil\break$d_{20}^+L_{aa}S\rho_{mm}
+240^+ L_{aa}S\tau
+d_{21}^+L_{ab}S\rho_{ab}
+d_{22}^+L_{ab}SR_{mabm}
+(195^+,105^-)L_{aa}L_{bb}E
+(30^+,150^-)L_{ab}L_{ab}E
+(195^+/6,105^-/6)L_{aa}L_{bb}\tau
+(5^+,25^-)L_{ab}L_{ab}\tau $\hfil\break$
+d_{23}^\pm L_{aa}L_{bb}\rho_{mm}
+d_{24}^\pm L_{ab}L_{ab}\rho_{mm}
+d_{25}^\pm L_{aa}L_{bc}\rho_{bc}
+d_{26}^\pm L_{aa}L_{bc}R_{mbcm}
+d_{27}^\pm L_{ab}L_{ac}\rho_{bc}
$\hfil\break$+d_{28}^\pm L_{ab}L_{ac}R_{mbcm}
+d_{29}^\pm L_{ab}L_{cd}R_{acbd}
+d_{30}^+ L_{aa}S^{3}
+d_{31}^+ L_{aa}L_{bb}S^{2}
+d_{32}^+ L_{ab}L_{ab}S^{2}
+d_{33}^+ L_{aa}L_{bb}L_{cc}S
+d_{34}^+ L_{aa}L_{bc}L_{bc}S
+d_{35}^+ L_{ab}L_{bc}L_{ac}S
+d_{36}^\pm L_{aa}L_{bb}L_{cc}L_{dd}
+$\hfil\break$d_{37}^\pm L_{aa}L_{bb}L_{cd}L_{cd}
+d_{38}^\pm L_{ab}L_{ab}L_{cd}L_{cd}
+d_{39}^\pm L_{aa}L_{bc}L_{cd}L_{db}
+d_{40}^\pm L_{ab}L_{bc}L_{cd}L_{da}
\}+$\hfil\break$F_{;m}\{
(90^+,-450^-)L_{aa}E
+d_{41}^\pm L_{aa}\rho_{mm}
+(15^+,-75^-)L_{aa}\tau
+d_{42}^+L_{aa}S^{2}
+d_{43}^\pm L_{aa:bb}
+d_{44}^\pm L_{ab:ab}
+d_{45}^\pm L_{ab}\rho_{ab}
+d_{46}^\pm L_{ab}R_{mabm}
+d_{47}^+L_{aa}L_{bb}S
+d_{48}^+L_{ab}L_{ab}S
+d_{49}^\pm L_{aa}L_{bb}L_{cc}
+d_{50}^\pm L_{aa}L_{bc}L_{bc}
+d_{51}^\pm L_{ab}L_{bc}L_{ac}
\}+F_{;mm}\{
d_{52}^+L_{aa}S
+d_{53}^\pm L_{aa}L_{bb}
+d_{54}^\pm L_{ab}L_{ab}
\}+$\hfil\break$F_{;mmm}
d_{55}^\pm L_{aa}\bigr\}[\partial M].$\smallbreak
\item{\rm 2)}  $ d \sb{ 43} \sp{ +}=151.875,$
$ d \sb{ 44} \sp{ +}=-11.25,$ and
$ d \sb{ 55} \sp{ +}=-30. $
\item{\rm 3)} $ d \sb{ 43} \sp{ -}=39.375,$
$ d \sb{ 44} \sp{ -}=-11.25,$ and
$ d \sb{ 55} \sp{ -}=-105. $
\endproclaim
\demo{Proof} We write down a basis for the space of invariants
using H.\ Weyl's theorem. We use Theorem 1.1 to evaluate the
coefficients of the expressions not involving the second
fundamental form $L$; we also Lemma 2.1 (2) to evaluate other
coefficients using suitable product formulas. The following invariants
can be expressed in terms of invariants already appearing in the list
above and are therefore omitted:
$FL_{aa:b}R_{cmcb},$
$FL_{ab:a}R_{cmcb},$
$FL_{ab}R_{acbc;m},$
$F\rho_{mm;m}L_{aa},$
$F\rho_{mm;m}S$, $F_{;m}\rho_{mm;m}$. We can also
replace `;' by `:' in some expressions at the cost
of introducing additional lower order terms involving
the second fundamental form. Thus, for example, by Lemma 6.1, we may express
$$\eqalign{&\rho_{mm;m}=(\tau_{;m}-R_{abba;m})/2=(\tau_{;m}-2R_{abbm;a})/2
\cr
=&(\tau_{;m}-2\rho_{am:a}+...)/2=(\tau_{;m}-2L_{bb:aa}+2L_{ab:ab}+...)/2\cr}$$
This completes the proof of 1).

\par
To prove 2) and 3), we suppress the coefficients of $  F$ and
$ F \sb{ ;mm}$ and we suppress terms of
length greater than 1 in the coefficients of $ F \sb{ ;m}$. This permits us
to express

$ a \sb{ 5}(F,D, {\cal B}  \sb{ S} \sp{
\pm })= \pm 5760 \sp {-1} (4 \pi ) \sp{
(m-1)/2}\ptr \big\{ F(*)+ $

\par\noindent
\PGb
$ +F \sb{ ;m} \{ (195/2 \sp{ +},60 \sp{
-}) \cdot  \tau  \sb{ ;m}+270S \sb{ :aa}+(630 \sp{
+},450 \sp{ -})E \sb{ ;m} $

\par\noindent
\PGb
$ +d \sb{ 43} \sp{  \pm }F \sb{ ;m}L \sb{
aa:bb}+d \sb{ 44} \sp{  \pm }F \sb{ ;m}L \sb{
ab:ab}+... \} +F \sb{ ;mm}(*) $

\par\noindent
\PGb
$ +F \sb{ ;mmm}(180S+d \sb{ 55} \sp{
\pm }L \sb{ aa})+45F \sb{ ;mm mm} \} [ \partial M].$
\et

 We have to be a bit careful
since we have not used linearly independent
monomials in the variational formulas in
\S6. Also, and equally importantly, we must
integrate by parts in computing certain integral
invariants. We apply Lemma 2.1 (7) to see that if
$ m=7,$ then
$$
0= \birdy a \sb{ 5}(e \sp{ -2 \epsilon f}F,e \sp{
-2 \epsilon f}D, {\cal B}  \sp{  \pm } \sb{
S( \epsilon )})  .
$$
The information in the tables below uses information from \S3
and from the appendix in \S6; the sum of each column multiplied by the
appropriate entry and coefficient is zero; this yields the equations
necessary to compute $ d \sb{ 43}^\pm,$
$ d \sb{ 44}^\pm,$ and
$ d \sb{ 55}^\pm$ and complete the proof. \hfill\qed\enddemo
\et\centerline{Neumann Boundary Conditions $m=7$}\et
\centerline{\vbox{
 {\offinterlineskip\halign{\vphantom{${}\sb{A}$}\vrule
 height 12pt\ #\hfill\ \vrule height
 12pt\ & \hfil#\ \vrule height 12pt\ &
 \hfil#\ \vrule height 12pt\ & \hfil#\ \vrule height 12pt\ &
 \hfil#\ \vrule height 12pt\ & \hfil#\ \vrule height 12pt\ &
 \hfil#\ \vrule height 12pt \cr \noalign{\hrule}
   Term&Coeff&$f_{;m}F_{;mmm}$&$f_{;a}F_{;m}L_{bb:a}$
 &$f_{;a}F_{;m}L_{ab:b}$&$f_{;mmm}F_{;m}$
 &$f_{;aam}F_{;m}$\cr\noalign{\hrule}
   $F_{;m}E_{;m}$&$630$&$0$&$0$&$0$&$2.5$&$2.5$\cr\noalign{\hrule}
   $F_{;m}\tau_{;m}$&$97.5$&$0$&0&$0$&$-12$&$-12$\cr\noalign{\hrule}
   $F_{;m}S_{:aa}$&$270$&$0$&$2.5$&$-5$&$0$&$2.5$\cr\noalign{\hrule}
   $F_{;mmmm}$&45&$-14$&$-4$&$-8$&$-9$&$-4$\cr\noalign{\hrule}
   $F_{;mmm}S$&$180$&$2.5$&$0$&0&$0$&0\cr\noalign{\hrule}
   $F_{;m}L_{aa:bb}$&$d_{43}^+$&$0$&$-4$&12&$0$&$-6$\cr\noalign{\hrule}
   $F_{;m}L_{ab:ab}$&$d_{44}^+$&$0$&$-2$&10&$0$&$-1$\cr\noalign{\hrule}
   $F_{;mmm}L_{aa}$&$d_{55}^+$&$-6$&$-3$&$0$&0&0\cr\noalign{\hrule}
   }}}}\et
\centerline{Dirichlet Boundary Conditions $m=7$}\et
 \centerline{\vbox{
 {\offinterlineskip\halign{\vphantom{${}\sb{A}$}\vrule
 height 12pt\ #\hfill\ \vrule height
 12pt\ & \hfil#\ \vrule height 12pt\ &
 \hfil#\ \vrule height 12pt\ & \hfil#\ \vrule height 12pt\ &
 \hfil#\ \vrule height 12pt\ & \hfil#\ \vrule height 12pt\ &
 \hfil#\ \vrule height 12pt \cr \noalign{\hrule}
   Term & Coeff &$f_{;m}F_{;mmm}$&$f_{;a}F_{;m}L_{bb:a}$
 &$f_{;a}F_{;m}L_{ab:b}$&$f_{;mmm}F_{;m}$
 &$f_{;aam}F_{;m}$\cr\noalign{\hrule}
   $F_{;m}E_{;m}$&450&$0$&0&$0$&2.5&2.5\cr\noalign{\hrule}
   $F_{;m}\tau_{;m}$&60&$0$&0&$0$&$-12$&$-12$\cr\noalign{\hrule}
   $F_{;mmmm}$&45&$-14$&$-4$&$-8$&$-9$&$-4$\cr\noalign{\hrule}
   $F_{;m}L_{aa:bb}$&$d_{43}^-$&$0$&$-4$&$12$&$0$&$-6$\cr\noalign{\hrule}
   $F_{;m}L_{ab:ab}$&$d_{44}^-$&$0$&$-2$&$10$&$0$&$-1$\cr\noalign{\hrule}
   $F_{;mmm}L_{aa}$&$d_{55}^-$&$-6$&$-3$&$0$&0&0\cr\noalign{\hrule}
   }}}}\et

\par\rm
We can now proceed with our study. We set
$ m=2,$ set
$F=1,$ and suppress all terms of length greater
than 2 to see
\PGa
$ a \sb{ 5}(1,D, {\cal B}  \sb{ S} \sp{
\pm })= \pm 5760 \sp {-1} (4 \pi ) \sp{
(m-1)/2}\ptr \{ (135/2) \tau  \sb{ ;mm} $

\par\noindent
\PGc
$ +270 \tau  \sb{ ;m}S-360S \sb{ :1}S \sb{
:1}+d \sb{ 1} \sp{  \pm }L \sb{ 11}E \sb{
;m}+e \sb{ 1} \sp{  \pm }L \sb{ 11:1}L \sb{
11:1} $

\par\noindent
\PGc
$ +e \sb{ 7} \sp{ +}L \sb{ 11:11}S+e_8^\pm L \sb{ 11}\tau+... \} . $
\proclaim{Lemma 5.2} $ e \sb{ 1} \sp{ +}=-53.4375,$
$ e \sb{ 7} \sp{ +}=270,$
$ e \sb{ 8} \sp{ +}=22.5, $
$ e \sp{ -} \sb{ 1}=2.8125,$
$ e \sb{ 8} \sp{ -}=-90. $\endproclaim
\demo{Proof} We consider the variation
$ D( \epsilon )=e \sp{ -2 \epsilon f}D$
with
$  {\cal B}  \sb{ S} \sp{  \pm }( \epsilon )$
defined appropriately and with
$ E(0)= \tau (0)=0.$ By Lemma 2.1 (4),
$$
-3a \sb{ 5}(f,D, {\cal B}  \sb{ S} \sp{
\pm })= \birdy a \sb{ 5}(1,e \sp{ -2 \epsilon
f}D, {\cal B}  \sp{  \pm } \sb{ S( \epsilon
)})  .
$$
We use this relation to evaluate
$ d \sb{ 2} \sp{  \pm },$
$ e \sb{ 1} \sp{  \pm },$ and
$ e \sb{ 7} \sp{ +}$ from the tables
given below; the variational formulas used
to create these tables are contained in \S6.
We also apply the results of \S3 to evaluate
certain of the coefficients.\hfill\qed
\et\centerline{Neumann Boundary Conditions $m=2$}\et
\centerline{\vbox{
 {\offinterlineskip\halign{\vphantom{${}\sb{\vrule height
  8pt}$}\vrule
 height 12pt\ #\hfill\ \vrule height
 12pt\ & \hfil#\ \vrule height 12pt\ &
 \hfil#\ \vrule height 12pt\ & \hfil#\ \vrule height 12pt\ &
 \hfil#\ \vrule height 12pt\ & \hfil#\ \vrule height 12pt\
  & \hfil#\ \vrule height 12pt\ & \hfil#\ \vrule height 12pt\cr
   \noalign{\hrule}
   Term & Coeff &$f_{;m}L_{11:11}$&$f_{;mmm}L_{11}$
 &$f_{;m}\tau_{;m}$&$f_{;m}S_{:11}$&$f_{;mmmm}$&$f_{;mmm}S$\cr
  \noalign{\hrule}
  $-3a_5(f,\cdot)$&&$-421.875$&$90$&
 $-292.5$&$-810$&$-135$&$-540$\cr
 \noalign{\hrule}
 $\tau_{;mm}$&$67.5$&$-4$&$2$&$-4$&$0$&$-2$&$0$\cr\noalign{\hrule}
 $\tau_{;m}S$&$270$&$0$&$0$&$0$&$-2$&0&$-2$\cr\noalign{\hrule}
 $L_{11:1}L_{11:1}$&$e_1^+$&2&$0$&$0$&$0$&0&$0$\cr\noalign{\hrule}
 $L_{11:11}S$&$e_7^+$&$0$&0&$0$&$-1$&0&$0$\cr\noalign{\hrule}
 $L_{11}\tau_{;m}$&$e_8^+$&$-2$&$-2$&$-1$&$0$&0&$0$\cr\noalign{\hrule}}}}}
\et\centerline{Dirichlet Boundary Conditions $m=2$}\et
\centerline{\vbox{
{\offinterlineskip\halign{\vphantom{${}\sb{\vrule 10pt}$}\vrule
 height 12pt\ #\hfill\ \vrule height
 12pt\ & \hfil#\ \vrule height 12pt\ &
 \hfil#\ \vrule height 12pt\ & \hfil#\ \vrule height 12pt\ &
 \hfil#\ \vrule height 12pt\ & \hfil#\ \vrule height 12pt\cr
   \noalign{\hrule}
   Term & Coeff &$f_{;m}L_{11:11}$&$f_{;mmm}L_{11}$
 &$f_{;m}\tau_{;m}$&$f_{;mmmm}$\cr
  \noalign{\hrule}
 $-3a_5(f,\cdot)$&&$-84.375$&$315$&$-180$&$-135$\cr
 \noalign{\hrule}
 $\tau_{;mm}$&$67.5$&$-4$&$2$&$-4$&$-2$\cr\noalign{\hrule}
 $L_{11:1}L_{11:1}$&$e_1^-$&2&$0$&$0$&0\cr\noalign{\hrule}
 $L_{11}\tau_{;m}$&$e_8^-$&$-2$&$-2$&$-1$&0\cr\noalign{\hrule}}}}}\enddemo
\head\S6 Appendix (Variational formulas)\endhead
\par\rm
In the appendix, we present some identities
used in the previous sections.  In Lemma
6.1, we recall the basic equations of structure
of \cite{\refBGa} and some other preliminary
results. In Lemma 6.2, we compute certain integral formulas
modulo lower order terms. In Lemma 6.3, we give
some variational formulas. Additional formulas useful in studying
manifolds with non totally geodesic boundaries
are available from the authors upon request
as are the proofs are entirely elementary
if somewhat tedious.
\proclaim{Lemma 6.1}\global\qctL=0
\pz $ \text{dvol} \sb{ m}=g \sp{ 1/2}dx \sb{
1} \cdot  \cdot  \cdot dx \sb{ m}, $ and
$  \Gamma  \sb{  \nu  \mu }{} \sp{  \sigma }= {\textstyle{1\over2}} g \sp{
\sigma  \gamma }( \partial  \sb{  \mu }g \sb{
\nu  \gamma }+ \partial  \sb{  \nu }g \sb{
\mu  \gamma }- \partial  \sb{  \gamma }g \sb{
\nu  \mu }). $
\pz If
$ D=-(g \sp{  \nu  \mu } \partial  \sb{
\nu } \partial  \sb{  \mu }+P \sp{  \sigma } \partial  \sb{
\sigma }+Q),$ then
$  \omega  \sb{  \nu }= {\textstyle{1\over2}} g \sb{
\nu  \mu }(P \sp{  \mu }+g \sp{  \sigma  \gamma } \Gamma  \sb{
\sigma  \gamma }{} \sp{  \mu }), $\hfil\break
$ E=Q-g \sp{  \nu  \mu }( \partial  \sb{
\mu } \omega  \sb{  \nu }+ \omega  \sb{
\nu } \omega  \sb{  \mu }- \omega  \sb{
\sigma } \Gamma  \sb{  \nu  \mu }{} \sp{
\sigma }),$
$ \Omega  \sb{  \nu  \mu }= \partial  \sb{
\nu } \omega  \sb{  \mu }- \partial  \sb{
\mu } \omega  \sb{  \nu }+ \omega  \sb{
\nu } \omega  \sb{  \mu }- \omega  \sb{
\mu } \omega  \sb{  \nu }. $
\pz
$ L \sb{  \alpha  \beta }=( \partial  \sb{
m}, \nabla \sb{ \partial \sb{ \alpha }} \partial  \sb{
\beta })= \Gamma  \sb{  \alpha  \beta }{} \sp{
m}=- {\textstyle{1\over2}} \sqrt{g \sp{
mm}}~ \partial  \sb{ m}g \sb{  \alpha  \beta }. $
\pz
$  \rho  \sb{ am}=L \sb{ bb:a}-L \sb{ ab:b}$, $ \rho  \sb{ am;m}=-R \sb{ am
mc;c}+ \rho  \sb{ mm;a}, $
$ R \sb{ ab ab;m}=2R \sb{ ab mb;a},$
$  \rho  \sb{ ij;j}= {\textstyle{1\over2}}  \tau  \sb{
;i},$
$  \rho  \sb{ am;a}= {\textstyle{1\over2}}  \tau  \sb{
;m}- \rho  \sb{ mm;m},$
$ 2 \rho  \sb{ mm;m}= \tau  \sb{ ;m}-R \sb{
ab ba;m},$
$ R \sb{ ab cm}=L \sb{ bc: a}-L \sb{
ac:b}, $\pz
$ f \sb{ ;a}=f \sb{ :a}, $
$ f \sb{ ;am}=f \sb{ ;ma}=f \sb{ ;m:a}+L \sb{
ab}f \sb{ :b},$
$ f \sb{ ;ab}=f \sb{ :ab}-L \sb{ ab}f \sb{
;m},$
$ f \sb{ ;ii}=f \sb{ ;mm}+f \sb{ :aa}-L \sb{
aa}f \sb{ ;m}, $
$ f \sb{ ;aa m}=f \sb{ ;m aa}- \rho  \sb{
am}f \sb{ ;a}- \rho  \sb{ mm}f \sb{ ;m}, $
$ f \sb{ ;ma a}=f \sb{ ;m:aa}+L \sb{
ab:a}f \sb{ :b}+2L \sb{ ac}f \sb{ :ac}-L \sb{
ac}L \sb{ ac}f \sb{ ;m}-L \sb{ aa}f \sb{
;mm}, $
$ f \sb{ ;ii m}=f \sb{ ;mm m}+f \sb{
;m:aa}+2L \sb{ ab:a}f \sb{ :b}-L \sb{ aa:b}f \sb{
:b}+2L \sb{ ac}f \sb{ :ac}-L \sb{ ac}L \sb{
ac}f \sb{ ;m } -L \sb{ aa}f \sb{ ;mm}- \rho  \sb{ mm}f \sb{
;m},$
$ f \sb{ ;aa mm}=f \sb{ ;mm aa}+2R \sb{
am mc;a}f \sb{ ;c}-2 \rho  \sb{ mm}f \sb{
;mm}+2R \sb{ am mb}f \sb{ ;ab}-f \sb{
;a} \rho  \sb{ mm;a } -2 \rho  \sb{ mb}f \sb{ ;bm}- \rho  \sb{
mm;m}f \sb{ ;m}. $
\endproclaim\et
We must integrate by parts in computing certain integral invariants.
The following formulas are useful; we suppress lower order terms of length
greater than 3.
\proclaim{Lemma 6.2}\global\qctL=0
\pz $  {\textstyle\int}  \sb{  \partial M}f \sb{
;m:aa}F \sb{ ;m}= {\textstyle\int}  \sb{
\partial M}(f \sb{ ;aa m}F \sb{ ;m}+F \sb{
;m}f \sb{ ;a}L \sb{ bb:a}-2F \sb{ ;m}f \sb{
;a}L \sb{ ab:b}) $.
\pz
$  {\textstyle\int}  \sb{  \partial M}f \sb{
;am}F \sb{ ;am}= {\textstyle\int}  \sb{
\partial M}(-f \sb{ ;aa m}F \sb{ ;m}-F \sb{ ;m}f \sb{ ;a}L \sb{
bb:a}+F \sb{ ;m}f \sb{ ;a}L \sb{ ab:b}+...)$.
\pz $  {\textstyle\int}  \sb{  \partial M}f \sb{
;a}F \sb{ ;mm a}= {\textstyle\int}  \sb{  \partial M}(-2f \sb{
;a}F \sb{ ;m}L \sb{ ab:b}+...)$.
\pz $  {\textstyle\int}  \sb{  \partial M}f \sb{
;aa mm}= {\textstyle\int}  \sb{  \partial M}(2f \sb{
;m}L \sb{ ab:ab}-f \sb{ ;mm m}L \sb{
aa}- {\textstyle{1\over2}}  \tau  \sb{ ;m}f \sb{
;m}+...) $.
\endproclaim

\par\rm
We consider the variation

\par\noindent
\PGa
$ D( \epsilon )=e \sp{ -2 \epsilon f}D,$
$ g( \epsilon )=e \sp{ 2 \epsilon f}g,$
$ F( \epsilon )=e \sp{ -2 \epsilon f}F.$
\et

\par\noindent
To keep the boundary conditions constant,
we set

\par\noindent
\PGa
$ S( \epsilon )=e \sp{ - \epsilon f}  \{  \omega  \sb{ m}(0)- \omega  \sb{
m}( \epsilon )+S \} .$ \et

\par\noindent
Let
$ e \sb{ i}$ be an orthonormal frame for
the tangent and cotangent bundles of
$  M$ with respect to the reference metric
$  g(0).$ Let
$ e \sb{ i}( \epsilon )=e \sp{ - \epsilon f}e \sb{
i}$ and
$ e \sp{ i}( \epsilon )=e \sp{  \epsilon f}e \sp{
i}$ be the corresponding frames for the metric
$ g( \epsilon ).$ We remark that contraction
and differentiation  do not commute;

\par\noindent
\PGa
$  \birdy ( \Phi  \sb{ ii})=( \birdy  \Phi ) \sb{
ii}-2f \Phi  \sb{ ii}$ \et

\par\noindent
for example. Although
$  \Gamma $ is not tensorial, its variation
is tensorial. Let

\par\noindent
\PGa
$ \dot  \Gamma  \sb{ ij}{} \sp{ k}:=( \birdy  \Gamma ) \sb{
ij}{} \sp{ k }$

\par\noindent
we keep the distinction between lower and
upper indices since

\par\noindent
\PGa
$ ( \birdy  \Gamma ) \sb{ ij}{} \sp{ k} \not= ( \birdy  \Gamma ) \sb{
ij k};$ (we shall not need
$ \dot  \Gamma  \sb{ ij k}). $

\proclaim{Lemma 6.3}\global\qctL=0
\pz
$ \dot  \Gamma  \sb{ ij}{} \sp{ k}:=$
$ ( \birdy  \Gamma ) \sb{ ij }{} \sp{ k}= \delta  \sb{
ik}f \sb{ ;j}+ \delta  \sb{ jk}f \sb{ ;i}- \delta  \sb{
ij}f \sb{ ;k}. $
\pz
$ ( \birdy  \Gamma ) \sb{ mm}{} \sp{ a}=-f \sb{
;a},$ and
$ ( \birdy  \Gamma ) \sb{ mm}{} \sp{ m}=f \sb{
;m}. $
\pz
$ ( \birdy L) \sb{ ab}=- \delta  \sb{ ab}f \sb{
;m}+fL \sb{ ab}. $
\pz
$  \birdy (S)=-fS+ {\textstyle{1\over2}} (m-2)f \sb{
;m} . $
\pz
$  \birdy (E)=-2fE+ {\textstyle{1\over2}} (m-2)f \sb{
;ii}. $
\pz
$ ( \birdy R) \sb{ ij k \ell }= \delta  \sb{
ik}f \sb{ ;j \ell }+ \delta  \sb{ j \ell }f \sb{
;ik}- \delta  \sb{ i \ell }f \sb{ ;jk}- \delta  \sb{
jk}f \sb{ ;il}+2fR \sb{ ij kl}. $
\pz
$ ( \birdy R) \sb{ ma bm}=- \delta  \sb{
ab}f \sb{ ;mm}-f \sb{ ;ab}+2fR \sb{ ma bm}. $
\pz
$ ( \birdy  \rho ) \sb{ ij}=(2-m)f \sb{
;ij}-g \sb{ ij}f \sb{ ;kk}. $
\pz
$  \birdy ( \rho  \sb{ mm})=-2f \rho  \sb{
mm}-f \sb{ ;aa}+(1-m)f \sb{ ;mm}. $
\pz
$  \birdy ( \tau )=-2f \tau +2(1-m)f \sb{
;ii}. $
\pz
$  \birdy (E \sb{ ;m})=-3fE \sb{ ;m}+ {\textstyle{1\over2}} (m-2)f \sb{
;ii m}-2f \sb{ ;m}E. $
\pz
$  \birdy ( \tau  \sb{ ;m})=-3f \tau  \sb{
;m}+2(1-m)f \sb{ ;ii m}-2f \sb{ ;m} \tau . $
\pz
$  \birdy ( \phi  \sb{ ;ii})=-2f \phi  \sb{
;ii}+( \birdy  \phi ) \sb{ ;ii}+(m-2)f \sb{
;i} \phi  \sb{ ;i}. $
\pz
$  \birdy ( \phi  \sb{ ;aa})=-2f \phi  \sb{
;aa}+( \birdy  \phi ) \sb{ ;aa}+(m-1)f \sb{
;m} \phi  \sb{ ;m}+(m-3)f \sb{ ;a} \phi  \sb{
;a}. $
\pz
$  \birdy ( \phi  \sb{ ;mm})=-2f \phi  \sb{
;mm}+( \birdy  \phi ) \sb{ ;mm}-f \sb{ ;m} \phi  \sb{
;m}+f \sb{ ;a} \phi  \sb{ ;a}. $
\pz
$  \birdy ( \rho  \sb{ mm;m})=-3f \rho  \sb{
mm;m}+(1-m)f \sb{ ;mm m}-f \sb{ ;aa m}+2f \sb{
;a} \rho  \sb{ am}-2f \sb{ ;m} \rho  \sb{
mm}.$
\pz
$  \birdy (F \sb{ ;mm})=$
$ -4fF \sb{ ;mm}-2f \sb{ ;mm}F-5f \sb{ ;m}F \sb{
;m}+f \sb{ ;a}F \sb{ ;a}. $
\pz
$  \birdy (F \sb{ ;mm m})=$
$ -5fF \sb{ ;mm m}-9f \sb{ ;m}F \sb{
;mm}+3f \sb{ ;a}F \sb{ ;am}-7f \sb{ ;mm}F \sb{
;m}+f \sb{ ;am}F \sb{ ;a } -2f \sb{ ;mm m}F. $
\pz
$  \birdy (E \sb{ ;aa})=-4fE \sb{ ;aa}+ {\textstyle{1\over2}} (m-2)f \sb{
;ii aa}-2f \sb{ ;aa}E+(m-7)f \sb{ ;a}E \sb{
;a}+(m-1)f \sb{ ;m}E \sb{ ;m}. $
\pz
$  \birdy (E \sb{ :aa})=-4fE \sb{ :aa}+ {\textstyle{1\over2}} (m-2)f \sb{
;ii:aa}-2f \sb{ :aa}E+(m-7)f \sb{ :a}E \sb{
:a}$
\pz
$  \birdy (E \sb{ ;mm})=-4fE \sb{ ;mm}+ {\textstyle{1\over2}} (m-2)f \sb{
;ii mm} -2f \sb{ ;mm}E+f \sb{ ;a}E \sb{ ;a}-5f \sb{
;m}E \sb{ ;m}. $
\pz
$  \birdy (E \sb{ ;ii})=-4fE \sb{ ;ii}+ {\textstyle{1\over2}} (m-2)f \sb{
;ii jj}-2f \sb{ ;ii}E+(m-6)f \sb{ ;a}E \sb{
;a}+(m-6)f \sb{ ;m}E \sb{ ;m}. $
\pz
$  \birdy ( \tau  \sb{ ;aa})=-4f \tau  \sb{
;aa}+2(1-m)f \sb{ ;ii aa}-2f \sb{ ;aa} \tau +(m-7)f \sb{
;a} \tau  \sb{ ;a} +(m-1)f \sb{ ;m} \tau  \sb{ ;m}. $
\pz
$  \birdy ( \tau  \sb{ :aa})=-4f \tau  \sb{
:aa}+2(1-m)f \sb{ ;ii:aa}-2f \sb{ :aa} \tau +(m-7)f \sb{
:a} \tau  \sb{ :a}.$
\pz
$  \birdy ( \tau  \sb{ ;mm})=-4f \tau  \sb{
;mm}+2(1-m)f \sb{ ;ii mm}-2f \sb{ ;mm} \tau +f \sb{
;a} \tau  \sb{ ;a} -5f \sb{ ;m} \tau  \sb{ ;m}. $
\pz
$  \birdy ( \tau  \sb{ ;ii})=-4f \tau  \sb{
;ii}+2(1-m)f \sb{ ;ii jj}-2f \sb{ ;ii} \tau +(m-6)f \sb{
;i} \tau  \sb{ ;i}. $
\pz
$  \birdy ( \rho  \sb{ mm;aa})=-4f \rho  \sb{
mm;aa}+(1-m)f \sb{ ;mm aa}-f \sb{ ;bb aa}-2f \sb{
;am} \rho  \sb{ am}-2f \sb{ ;aa} \rho  \sb{ mm}-4f \sb{ ;m} \rho  \sb{
am;a}+(m-7)f \sb{ ;a} \rho  \sb{ mm;a}+(m-1)f \sb{
;m} \rho  \sb{ mm;m}. $
\pz
$  \birdy ( \rho  \sb{ mm:aa})=-4f \rho  \sb{
mm:aa}-f \sb{ ;aa:bb}+(1-m)f \sb{ ;mm:aa}-2f \sb{ :aa} \rho  \sb{mm}+$
\hfil\break $(m-7)f\sb{ :a} \rho  \sb{ mm:a}. $
\pz
$  \birdy ( \rho  \sb{ mm;mm})=-4f \rho  \sb{
mm;mm}-f \sb{ ;aa mm}+(1-m)f \sb{ ;mm mm}-2f \sb{ ;mm} \rho  \sb{ mm}+f
\sb{ ;a} \rho  \sb{ mm;a}+4f \sb{ ;a} \rho  \sb{ am;m}+2f \sb{
;am} \rho  \sb{ am}-5f \sb{ ;m} \rho  \sb{
mm;m }$
\pz
$  \birdy ( \tau E)=-4f \tau E+ {\textstyle{1\over2}} (m-2)f \sb{
;ii} \tau +2(1-m)f \sb{ ;ii}E.$
\pz
$  \birdy (E \sp{ 2})=-4fE \sp{ 2}+(m-2)f \sb{
;ii}E.$
\pz
$  \birdy ( \tau  \sp{ 2})=-4f \tau  \sp{
2}+4(1-m)f \sb{ ;ii} \tau .$
\pz
$  \birdy ( \rho  \sp{ 2})=-4 f\rho  \sp{
2}-2f \sb{ ;ii} \tau +2(2-m)f \sb{ ;ij} \rho  \sb{
ij}.$
\pz
$  \birdy (R \sp{ 2})=-4fR \sp{ 2}-8f \sb{
;ij} \rho  \sb{ ij}.$
\pz
$  \birdy ( \Omega  \sp{ 2})=-4f \Omega  \sp{
2}.$
\pz
$  \birdy ( \rho  \sb{ mm}E)=-4f \rho  \sb{
mm}E-f \sb{ ;aa}E+(1-m)f \sb{ ;mm}E+ {\textstyle{1\over2}} (m-2)f \sb{ ;ii}
\rho  \sb{ mm}.$
\pz
$  \birdy ( \rho  \sb{ mm} \tau )=-4f \rho  \sb{
mm} \tau -f \sb{ ;aa} \tau +(1-m)f \sb{
;mm} \tau  +2(1-m)f \sb{ ;ii} \rho  \sb{ mm}. $
\pz
$  \birdy (R \sb{ am mb} \rho  \sb{ ab})=-4fR \sb{
am mb} \rho  \sb{ ab}-f \sb{ ;ab} \rho  \sb{
ab}-f \sb{ ;mm} \tau -f \sb{ ;aa} \rho  \sb{
mm}+$\hfil\break$(2-m)R \sb{ am mb}f \sb{ ;ab}. $
\pz
$  \birdy ( \rho  \sb{ mm} \sp{ 2})=-4 f\rho  \sb{
mm} \sp{ 2}-2f \sb{ ;aa} \rho  \sb{ mm}+2(1-m)f \sb{
;mm} \rho  \sb{ mm}.$
\pz
$  \birdy (R \sb{ am mb}R \sb{ am mb})=-4fR \sb{
am mb}R \sb{ am mb}-2f \sb{ ;ab}R \sb{
am mb} -2f \sb{ ;mm} \rho  \sb{ mm}. $
\pz$\birdy ( \Omega  \sb{ am} \Omega  \sb{
am})=-4f \Omega  \sb{ am} \Omega  \sb{ am}.$
\pz
$  \birdy (E \sb{ ;m}S)=-4fE \sb{ ;m}S+ {\textstyle{1\over2}} (m-2)f \sb{
;ii m}S-2f \sb{ ;m}ES+ {\textstyle{1\over2}} (m-2)f \sb{ ;m}E \sb{ ;m}. $
\pz
$  \birdy (ES \sp{ 2})=-4fE S \sp{ 2}+ {\textstyle{1\over2}} (m-2)f \sb{
;ii}S \sp{ 2}+(m-2)f \sb{ ;m}ES.$
\pz
$  \birdy (S \sp{ 4})=-4fS \sp{ 4}+2(m-2)f \sb{
;m}S \sp{ 3}.$
\pz
$  \birdy ( \tau  \sb{ ;m}S)=-4f \tau  \sb{
;m}S+2(1-m)f \sb{ ;ii m}S+ {\textstyle{1\over2}} (m-2)f \sb{
;m}  \tau  \sb{ ;m} -2f \sb{ ;m} \tau S. $
\pz
$  \birdy ( \rho  \sb{ mm}S \sp{ 2})=-4f \rho  \sb{
mm}S \sp{ 2}-f \sb{ ;aa}S \sp{ 2}+(1-m)f \sb{
;mm}S \sp{ 2}+(m-2)f \sb{ ;m} \rho  \sb{ mm}S. $
\pz
$  \birdy ( \tau S \sp{ 2})=-4f \tau S \sp{
2}+2(1-m)f \sb{ ;ii}S \sp{ 2}+(m-2)f \sb{
;m} \tau S.$
\pz
$  \birdy (S \sb{ :aa}S)=-4fS \sb{ :aa}S-f \sb{
:aa}S \sp{ 2}+(m-5)f \sb{ ;a}S \sb{ :a}S+ {\textstyle{1\over2}} (m-2)  f
\sb{ ;m :aa}S+$\hfil\break$ {\textstyle{1\over2}} (m-2)S \sb{
:aa}f \sb{ ;m} . $
\pz
$  \birdy (S \sb{ :a}S \sb{ :a})=-4fS \sb{
:a}S \sb{ :a}-2f \sb{ ;a}S \sb{ :a}S+(m-2)   f \sb{ ;m :a}S \sb{ :a}.$
\pz
$  \birdy (F \sb{ ;m}E \sb{ ;m})=-6fF \sb{
;m}E \sb{ ;m}-2f \sb{ ;m}FE \sb{ ;m}
+ {\textstyle{1\over2}} (m-2)f \sb{ ;ii m}F \sb{
;m}-2f \sb{ ;m}F \sb{ ;m}E.$
\pz
$  \birdy (F \sb{ ;m}ES)=-6fF \sb{ ;m}ES-2f \sb{
;m}FE S+ {\textstyle{1\over2}} (m-2)f \sb{ ;ii}F \sb{
;m}S+$\hfil\break$ {\textstyle{1\over2}} (m-2)f \sb{
;m}F \sb{ ;m}E. $
\pz
$  \birdy (F \sb{ ;m}S \sp{ 3})=-6fF \sb{
;m}S \sp{ 3}-2f \sb{ ;m}FS \sp{ 3}+ {\textstyle{3\over{2}}} (m-2)f \sb{
;m}F \sb{ ;m}S \sp{ 2}.$
\pz
$  \birdy (F \sb{ ;m} \tau  \sb{ ;m})=-6fF \sb{
;m} \tau  \sb{ ;m}-2f \sb{ ;m}F \tau  \sb{
;m} +2(1-m)f \sb{ ;ii m}F \sb{ ;m}-2f \sb{
;m}F \sb{ ;m} \tau . $
\pz
$  \birdy (F \sb{ ;m} \tau S)=-6fF \sb{
;m} \tau S-2f \sb{ ;m}F \tau S
+2(1-m)f \sb{ ;ii}F \sb{ ;m}S+ {\textstyle{1\over2}} (m-2)f \sb{
;m}F \sb{ ;m} \tau  . $
\pz
$  \birdy (F \sb{ ;m} \rho  \sb{ mm}S)=-6fF \sb{
;m} \rho  \sb{ mm}S-2f \sb{ ;m}F \rho  \sb{
mm}S-f \sb{ ;aa}F \sb{ ;m}S+$\hfil\break$(1-m)f \sb{ ;mm}F \sb{
;m}S+ {\textstyle{1\over2}} (m-2)f \sb{
;m}F \sb{ ;m} \rho  \sb{ mm}. $
\pz
$  \birdy (F \sb{ ;m}S \sb{ :aa})=-6fF \sb{
;m}S \sb{ :aa}-2f \sb{ ;m}FS \sb{ :aa}-f \sb{ :aa}F \sb{ ;m}S+(m-5)f \sb{
;a}F \sb{ ;m}S \sb{ :a}+ {\textstyle{1\over2}} (m-2)F \sb{
;m} f \sb{ ;m:aa}. $
\pz
$  \birdy (F \sb{ ;mm}E)=-6fF \sb{ ;mm}E+f \sb{
;a}F \sb{ ;a}E-2f \sb{ ;mm}FE
 -5f \sb{ ;m}F \sb{ ;m}E+$\hfil\break$ {\textstyle{1\over2}} (m-2)f \sb{
;ii}F \sb{ ;mm }.$
\pz
$  \birdy (F \sb{ ;mm}S \sp{ 2})=$
$ -6fF \sb{ ;mm}S \sp{ 2}+f \sb{ ;a}F \sb{
;a}S \sp{ 2} -2f \sb{ ;mm}FS \sp{ 2}-5f \sb{ ;m}F \sb{
;m}S \sp{ 2}+$\hfil\break$(m-2)f \sb{ ;m}F \sb{ ;mm}S. $
\pz
$  \birdy (F \sb{ ;mm} \tau )=-6fF \sb{
;mm} \tau +f \sb{ ;a}F \sb{ ;a} \tau -2f \sb{
;mm}F \tau -5f \sb{ ;m}F \sb{ ;m} \tau
 +2(1-m)f \sb{ ;ii}F \sb{ ;mm}. $
\pz
$  \birdy (F \sb{ ;mm} \rho  \sb{ mm})=-6fF \sb{
;mm} \rho  \sb{ mm}+f \sb{ ;a}F \sb{ ;a} \rho  \sb{
mm} -2f \sb{ ;mm}F \rho  \sb{ mm}-5f \sb{
;m}F \sb{ ;m} \rho  \sb{ mm}-f \sb{ ;aa}F \sb{
;mm}+(1-m)f \sb{ ;mm}F \sb{ ;mm}. $
\pz
$  \birdy (F \sb{ ;mm m}S)=-6fF \sb{
;mm m}S+3f \sb{ ;a}F \sb{ ;am}S-7f \sb{
;mm}F \sb{ ;m}S -2f \sb{ ;mm m}FS-$\hfil\break$9f \sb{ ;m}F \sb{
;mm}S+f \sb{ ;am}F \sb{ ;a}S+ {\textstyle{1\over2}} (m-2)f \sb{
;m}F \sb{ ;mm m}. $
\pz
$  \birdy (F \sb{ ;mm mm})=-6fF \sb{
;mm mm}-9f \sb{ ;mm m}F \sb{ ;m}+f \sb{
;mm a}F \sb{ ;a}
 -16f \sb{ ;mm}F \sb{ ;mm}+4f \sb{ ;am}F \sb{
;am}-14f \sb{ ;m}F \sb{ ;m mm}+6f \sb{
;a}F \sb{ ;mm a }
+6f \sb{ ;a}R \sb{ am mb}F \sb{ ;b}-2f \sb{
;mm mm}F.$
\pz
$  \birdy (L \sb{ aa} \tau  \sb{ :m})=-4fL \sb{
aa} \tau  \sb{ ;m}+(1-m)f \sb{ ;m} \tau  \sb{
;m} +2(1-m)f \sb{ ;ii    m}L \sb{ aa}-2f \sb{
;m}L \sb{ aa} \tau . $
\pz
$  \birdy (L \sb{ aa:bb}S)=-4fL \sb{ aa:bb}S \sp{
}+(1-m)f \sb{ ;m:aa}S+(m-5)f \sb{ ;b}L \sb{
aa:b}S-f \sb{ :bb}L \sb{ aa}S + {\textstyle{1\over2}} (m-2)f \sb{ ;m}L
\sb{ aa:bb}. $
\pz
$   \birdy (L \sb{ ab:c}L \sb{ ab:c})=-4fL \sb{
ab:c}L \sb{ ab:c}-2f \sb{ ;m:c}L \sb{ aa:c}+4f \sb{
:e}L \sb{ eb}L \sb{ ab:a}
 -2f \sb{ ;c}L \sb{ ab}L \sb{ ab:c}-4f \sb{
;a}L \sb{ cb}L \sb{ ab:c}. $
\pz
$  \birdy (F \sb{ ;m}L \sb{ aa:bb})=-6fF \sb{
;m}L \sb{ aa:bb}-2f \sb{ ;m}FL \sb{ aa:bb}
+(1-m)f \sb{ ;m:bb}F \sb{ :m}+$\hfil\break$(m-5)f \sb{
;b}F \sb{ ;m}L \sb{ aa:b}-f \sb{ :aa}F \sb{
;m}L \sb{ bb}. $
\pz
$  \birdy (F \sb{ ;m}L \sb{ ab:ab})=-6fF \sb{
;m}L \sb{ ab:ab}-2f \sb{ ;m}FL \sb{ ab:ab}
-f \sb{ ;m:bb}F \sb{ ;m}+(m-2)f \sb{ :eb}F \sb{
;m}L \sb{ eb}$\hfil\break$+2(m-3)f \sb{ :e}F \sb{ ;m}L \sb{
ae:a }
 -f \sb{ :bb}F \sb{ ;m}L \sb{ aa}-f \sb{
:e}F \sb{ ;m}L \sb{ aa:e }$.
\pz
$  \birdy (F \sb{ ;mm    m}L \sb{ aa})=-6fF \sb{
;mm    m}L \sb{ aa}-9f \sb{ ;m}F \sb{ ;mm}L \sb{
aa}+3f \sb{ ;b}F \sb{ ;bm}L \sb{ aa}
 -$\hfil\break$7f \sb{ ;mm}F \sb{ ;m}L \sb{ aa}+f \sb{
;am}F \sb{ ;a}L \sb{ bb}-2f \sb{ ;mm    m}FL \sb{
aa}+(1-m)f \sb{ ;m}F \sb{ ;mm    m}.$

\endproclaim
\proclaim{Remark}\rm Branson and Gilkey \cite{\refBGa, Theorem 7.2} also
computed
$a_n$ for mixed boundary conditions. Their formula for $a_4$ gave incorrectly
the
values $\beta_3=-42$ and $\beta_4=6$. Vassilevich \cite{\refVa} showed the
correct
values were
$\beta_3=-12$ and $\beta_4=-24$. The error in Branson-Gilkey arose from
incorrectly applying the variational formula contained in Lemma 6.3 (4),
which {\bf is} valid for Neumann boundary conditions, to the more general
context
of mixed boundary conditions. If one were to consider mixed boundary
conditions,
the relevant variational formula for $S$ would become
$\birdy(S)=-fS+(m-2)f_{;m}\Pi^+/2$ and the other variational formulas involving
$S$ in Lemma 6.3 would then need to be changed accordingly. We wish to caution
the reader concerning this point.\endproclaim

\par\noindent
Thomas P. Branson: Department of mathematics,
University of Iowa, Iowa City IA 52242 USA.
email: branson\@blue.weeg.uiowa.edu. Research partially supported by an
international travel grant of the NSF (USA).\et

\par\noindent
Peter B. Gilkey: Department of mathematics,
University of Oregon, Eugene Oregon 97403
USA. email: gilkey\@math.uoregon.edu. Research
partially supported by IHES (France) and
the NSF (USA).\et

\par\noindent
Dmitri V Vassilevich: Department of Theoretical
Physics, St. Petersburg University, 198904
St. Petersburg Russia. email: dvassil\@sph.spb.su. Research
partially supported by GRACENAS (Russia) \vfill
29 March 1995 version 35
\enddocument